\documentclass[11pt,draftclsnofoot,peerreviewca,onecolumn,letterpaper]{IEEEtran}
\usepackage{amsfonts,amsmath,amssymb}
\usepackage{graphicx,color,epsfig,rotating}
\usepackage{latexsym}
\usepackage{subfigure}
\usepackage{cite}
\usepackage{geometry}

\long\def\comment#1{}

\newfont{\bbb}{msbm10 scaled 700}

\newfont{\bb}{msbm10 scaled 1100}
\newcommand{\CC}{\mbox{\bb C}}

\newcommand{\RR}{\mbox{\bb R}}

\newcommand{\EE}{\mbox{\bb E}}

\newcommand{\av}{{\bf a}}
\newcommand{\bv}{{\bf b}}
\newcommand{\cv}{{\bf c}}
\newcommand{\dv}{{\bf d}}

\newcommand{\gv}{{\bf g}}
\newcommand{\hv}{{\bf h}}

\newcommand{\pv}{{\bf p}}
\newcommand{\qv}{{\bf q}}
\newcommand{\rv}{{\bf r}}
\newcommand{\sv}{{\bf s}}
\newcommand{\tv}{{\bf t}}
\newcommand{\uv}{{\bf u}}
\newcommand{\wv}{{\bf w}}
\newcommand{\vv}{{\bf v}}
\newcommand{\xv}{{\bf x}}
\newcommand{\yv}{{\bf y}}
\newcommand{\zv}{{\bf z}}
\newcommand{\zerov}{{\bf 0}}
\newcommand{\onev}{{\bf 1}}

\newcommand{\Am}{{\bf A}}
\newcommand{\Bm}{{\bf B}}
\newcommand{\Cm}{{\bf C}}
\newcommand{\Dm}{{\bf D}}

\newcommand{\Gm}{{\bf G}}
\newcommand{\Hm}{{\bf H}}
\newcommand{\Id}{{\bf I}}

\newcommand{\Rm}{{\bf R}}
\newcommand{\Sm}{{\bf S}}
\newcommand{\Tm}{{\bf T}}
\newcommand{\Um}{{\bf U}}

\newcommand{\Vm}{{\bf V}}

\newcommand{\Cc}{{\cal C}}

\newcommand{\Gc}{{\cal G}}

\newcommand{\Lc}{{\cal L}}

\newcommand{\Nc}{{\cal N}}

\newcommand{\Rc}{{\cal R}}
\newcommand{\Sc}{{\cal S}}

\newcommand{\Uc}{{\cal U}}

\newcommand{\gammav}{\hbox{\boldmath$\gamma$}}

\newcommand{\lambdav}{\hbox{\boldmath$\lambda$}}

\newcommand{\Sigmam}{\hbox{\boldmath$\Sigma$}}
\newcommand{\Phim}{\hbox{\boldmath$\Phi$}}

\newcommand{\Psim}{\hbox{\boldmath$\Psi$}}

\newcommand{\Xim}{\hbox{\boldmath$\Xi$}}

\renewcommand{\det}{{\hbox{det}}}
\newcommand{\trace}{{\hbox{tr}}}

\newcommand{\SINR}{{\sf SINR}}

\renewcommand{\Re}{{\rm Re}}
\renewcommand{\Im}{{\rm Im}}
\newcommand{\eqdef}{\stackrel{\Delta}{=}}

\newcommand{\herm}{{\sf H}}

\newcommand{\transp}{{\sf T}}

\newtheorem{theorem}{Theorem}

\newcommand{\xsf}{{\bold{\sf x}}}

\begin{document}

\title{Multiuser MISO Transmitter Optimization for Inter-Cell Interference Mitigation}

\author{Hoon Huh,~\IEEEmembership{Student Member,~IEEE}, Haralabos C. Papadopoulos,~\IEEEmembership{Member,~IEEE},
\\ and Giuseppe Caire*,~\IEEEmembership{Fellow,~IEEE}
\thanks{H. Huh and G. Caire are with the Department of Electrical Engineering, University of Southern California, Los Angeles, CA 90089, USA. (e-mail: hhuh, caire@usc.edu)}
\thanks{H. C. Papadopoulos is with DoCoMo USA Labs, Palo Alto, CA 94304, USA (e-mail: hpapadopoulos@docomolabs-usa.com)}
\thanks{The material in this paper was presented in part at the IEEE International Symposium on Information Theory, ISIT 2009 in Seoul, Korea.}}

\thispagestyle{empty}

\maketitle

\begin{abstract}
The transmitter optimization (i.e., steering vectors and power allocation) for a MISO Broadcast Channel (MISO-BC) subject to general linear constraints is considered. Such constraints include, as special cases,  the sum power, the per-antenna or per-group-of-antennas
power, and ``forbidden interference direction'' constraints.
We consider both the optimal dirty-paper coding and the simple suboptimal linear zero-forcing beamforming strategies,
and provide numerically efficient algorithms that solve the problem in its most general form.
As an application, we consider a multi-cell scenario with partial cell cooperation,
where each cell optimizes its precoder by taking into account interference constraints on specific users
in adjacent cells.  The effectiveness of the proposed methods is evaluated in a simple system scenario
including two adjacent cells, under different fairness criteria that emphasize the bottleneck role of users near
the cell ``boundary''.  Our results show that ``active'' Inter-Cell Interference (ICI) mitigation
outperforms the conventional ``static'' ICI mitigation based on {\em fractional} frequency reuse.
\end{abstract}


\begin{center}
\bfseries EDICS Categories: MSP-CODR, MSP-APPL
\end{center}

\newpage
\setcounter{page}{1}

\section{Problem setup and summary of the results} \label{sec:intro}

The next generation of wireless communication systems
(e.g., 802.16m \cite{IEEE80216m}, LTE-Advanced \cite{LTE-Advanced})
is expected to capitalize on the large spectral efficiency gains
promised by multiuser MIMO communications.

The fundamental information-theoretic model for the downlink of a cellular system comprising
one base-station equipped with $M$ transmit antennas and $K$ user terminals, each one with one or more receiving antennas,
is the well-known Gaussian MIMO Broadcast Channel (MIMO-BC) \cite{Caire-Shamai-TIT03, Viswanath-Tse-TIT03, Vishwanath-Jindal-Goldsmith-TIT03, Yu-Cioffi-TIT04, Weingarten-Steinberg-Shamai-TIT04}.
In this work we restrict to the case of single-antenna receivers, commonly referred to as the Multi-Input Single-Output Broadcast Channel (MISO-BC).
One channel use of the MISO-BC is described by
\begin{equation} \label{eq:general-down}
y_k = \hv_k^\herm \xv + z_k, \;\;\; k = 1,\ldots, K
\end{equation}
where $\hv_k \in \CC^M$ denotes the channel vector of user $k$, $\xv \in \CC^M$ is the transmitted signal vector
and $z_k \sim \Cc\Nc(0,1)$ is a complex circularly symmetric additive white Gaussian noise (AWGN).

Let $\Sc$ denote a {\em compact} set of $M \times M$ covariance matrices. The capacity region of the MISO-BC
(\ref{eq:general-down}) subject to the input constraint $\EE[ \xv \xv^\herm] \eqdef \Sigmam_x \in \Sc$
is given by the set of rate vectors $\Rm \in \RR_+^K$ \cite{Weingarten-Steinberg-Shamai-TIT04}
\begin{eqnarray} \label{eq:cap-region}
\Cc_{\rm bc}(\Sc; \Hm) & = & \mbox{coh} \; \bigcup_{\sum_{k=1}^K \vv_k \vv_k^\herm q_k  \in \Sc} \;\; \bigcup_{\pi} \nonumber \\
&& \left \{ R_{\pi_k} \leq \log \left (1 + \frac{|\hv_{\pi_k}^\herm \vv_{\pi_k}|^2 q_{\pi_k}}{1 +
\sum_{j=k+1}^K |\hv_{\pi_k}^\herm \vv_{\pi_j}|^2 q_{\pi_j}} \right ), \;\; \forall \; k \right \} \nonumber \\
&&
\end{eqnarray}
where the $M \times K$ channel matrix $\Hm = [\hv_1,\ldots, \hv_K]$ collects all user channel vectors.
The capacity region $\Cc_{\rm bc}(\Sc; \Hm)$ is achieved by {\em Gaussian Dirty-Paper Coding} (DPC), where the permutation
$\pi = (\pi_1,\ldots, \pi_K)$  of the user indices $\{1,\ldots, K\}$ indicates the {\em successive encoding order} where user $\pi_1$ is encoded first and user $\pi_K$ is encoded last. The transmit covariance matrix is given by $\Sigmam_x = \sum_{k=1}^K \vv_k \vv_k^\herm q_k$ and it is defined
by the unit-norm ``steering vectors'' $\{\vv_k\}$ and by the users transmit powers $\{q_k\}$.

The transmitter parameters $\{\vv_k\}$, $\{q_k\}$, $\pi$ achieving the boundary of $\Cc_{\rm bc}(\Sc; \Hm)$ can be determined by solving
the Weighted Rate Sum Maximization (WSRM) problem
\begin{eqnarray} \label{eq:wsrm}
\mbox{maximize} & & \sum_{k=1}^K W_k R_k \nonumber \\
\mbox{subject to} & & \Rm \in \Cc_{\rm bc}(\Sc; \Hm)
\end{eqnarray}
for some suitable choice of the nonnegative weights $\{W_k\}$. Although a direct solution of (\ref{eq:wsrm}) is generally difficult,
for the special case where the constraint set $\Sc$ is defined by {\em linear inequalities}
\begin{equation} \label{eq:lin-constr}
\trace \left (\Sigmam_x \Phim_\ell \right ) \leq \gamma_\ell,  \;\;\; \ell = 1,\ldots, L,
\end{equation}
where $\{\Phim_\ell\}$ are positive semidefinite symmetric matrices and $\{\gamma_\ell\}$ are non-negative coefficients,
the solution of (\ref{eq:wsrm}) can be computed efficiently by solving a sequence of convex problems.

By the Heine-Borel theorem, the compactness of $\Sc$ implies that $\Sc$ is {\em bounded} with respect to the Frobenius norm. Hence, without loss of generality, we can always include an additional sum-power constraint $\trace(\Sigmam_x) \leq P$ for some sufficiently large $P$, without modifying the problem. This corresponds to including an additional constraint with $\gamma_0 = P$ and $\Phim_0 = \Id$ in the set of constraints (\ref{eq:lin-constr}).
Important special cases of the constraint (\ref{eq:lin-constr}), beyond the sum-power constraint already discussed \cite{Caire-Shamai-TIT03, Viswanath-Tse-TIT03, Vishwanath-Jindal-Goldsmith-TIT03, Yu-Cioffi-TIT04}, include the per-antenna power constraint and the per-group of antennas constraint \cite{Yu-Lan-TSP07}.
For rank-1 $\Phim_\ell = \cv_\ell \cv_\ell^\herm$, we have a general ``interference'' constraint where the vector $\cv_\ell$ denotes a forbidden
direction along which the transmit power must be not larger than $\gamma_\ell$ \cite{Zhang-Poor-etal-ISIT09}.

Linear beamforming is a suboptimal precoding strategy that provides a low-complexity alternative to DPC.
When combined with Gaussian random coding, the following region is achievable
\begin{eqnarray} \label{eq:bf-region}
\Rc_{\rm bc}^{\rm bf}(\Sc; \Hm) & = & \mbox{coh} \; \bigcup_{\sum_{k=1}^K \vv_k \vv_k^\herm q_k  \in \Sc}
\left \{ R_{k} \leq \log \left (1 + \frac{|\hv_{k}^\herm \vv_{k}|^2 q_{k}}{1 +
\sum_{j\neq k} |\hv_{k}^\herm \vv_{j}|^2 q_{j}} \right ), \;\; \forall \; k \right \} \nonumber \\
&&
\end{eqnarray}
The optimization of the transmitter parameters $\{\vv_k\}, \{q_k\}$ is generally more difficult than with DPC since the WSRM problem
with linear beamforming has no general convex programming equivalent.
In this work we focus on the popular {\em Zero-Forcing Beamforming} (ZFBF) for at least three good reasons:
1) in the regime of high SNR and/or if combined with user selection \cite{Dimic-Sidiropoulos-TSP05, Yoo-Goldsmith-JSAC06} it yields near-optimal performance;
2) its performance is well-understood and extensively analyzed also in the case of  non-ideal channel state information \cite{Ding-Love-Zoltowski-TSP07, Jose-Vishwanath-Allerton08, Caire-Jindal-Kobayashi-Ravindran-ISIT07}, and therefore it is practically relevant for applications;
3) it lends itself to a computationally efficient solution.

The WSRM problem with ZFBF and general linear constraints is given by
\begin{eqnarray} \label{eq:wsrm-zfbf}
{\rm maximize} &  & \sum_{k=1}^K W_k \log\left (1 +  |\hv_k^\herm \vv_k|^2 q_k \right ) \nonumber \\
\mbox{subject to} &  & \hv_j^\herm \vv_k = 0 \;\;\;\; \forall j \neq k \nonumber \\
& & \trace \left (\Sigmam_x \Phim_\ell \right ) \leq \gamma_\ell, \;\;\; \forall \ell
\end{eqnarray}
Without loss of fundamental generality we consider the case where $\Hm$ has rank $K$ with $K \leq M$, otherwise the problem is infeasible.
If $K>M$, some user selection algorithm such as those proposed in \cite{Dimic-Sidiropoulos-TSP05, Yoo-Goldsmith-JSAC06}
takes care of selecting an ``active user subset'' of size not larger than $M$.  In fact in this paper, \cite{Dimic-Sidiropoulos-TSP05} is combined with the algorithm we develop to solve (\ref{eq:wsrm-zfbf}).


In the rest of this paper we consider problems (\ref{eq:wsrm}) and (\ref{eq:wsrm-zfbf}) subject to general linear constraints.
Although these problems have been addressed in a number of recent papers, a thorough comparison of the various algorithms
and a unified presentation has been missing. As far as problem (\ref{eq:wsrm}) is concerned, we show the equivalence of the
``SINR-duality'' \cite{Zhang-Poor-etal-ISIT09}  and the ``min-max duality'' \cite{Yu-Lan-TSP07} approaches, and give the computation details of the infeasible
start Newton iteration algorithm for general linear constraints (only sketched in \cite{Yu-Lan-TSP07} for the case of per-antenna power constraint).
Also, we show through simulation examples that this algorithm converges much faster and it is generally less complex than the inner-outer
iterative algorithm based on Lagrangian duality and sub-gradient search advocated in \cite{Zhang-Poor-etal-ISIT09}.
As far as problem (\ref{eq:wsrm-zfbf}), we generalize the convex relaxation approach of \cite{Wiesel-Eldar-Shamai-TSP08} to the case of
linear constraints and arbitrary rate weights (not immediately obvious from \cite{Wiesel-Eldar-Shamai-TSP08}).
We solve the convex relaxation problem using a novel gradient descent algorithm with logarithmic barrier.
Also, we propose a novel two-step iterative algorithm that updates directly the steering vectors of the ZF precoder,
building on the form of generalized inverses.  This new method is significantly more computationally efficient than the approach based on convex relaxation since it
avoids the dimensionality expansion of convex relaxation.  Finally, we use the proposed optimization algorithms in a multi-cell ICI mitigation scheme,
where each cell optimizes its transmit covariance matrix by taking into account an interference constraint on the edge users of the adjacent cell.
In a simple linear two-cell scenario, it is showed by Monte Carlo simulation that this approach can effectively improve the rate of the edge users
and achieves generally better rates than simpler conventional schemes such as FFR.

\section{WSRM algorithms for DPC} \label{sec:dpc}



Consider the problem (\ref{eq:wsrm}) where $\Sc$ is a compact convex set defined by the linear constraints
(\ref{eq:lin-constr}), including a (possibly irrelevant) sum-power constraint, as said before.
Without loss of generality, assume $W_1 \geq \cdots \geq W_K > 0$.
In \cite{Zhang-Poor-etal-ISIT09}, using a technique called ``Signal-to-Interference plus Noise Ratio (SINR) duality'',
the following fundamental results are proved.  Define the ``dual MAC'' corresponding to (\ref{eq:general-down}) as the multiple-access Gaussian channel
\begin{equation} \label{eq:dual-mac}
\yv = \sum_{k=1}^K \hv_k x_k + \zv
\end{equation}
where $\yv, \zv \in \CC^M$, $\zv \sim \Cc\Nc(\zerov, \Sigmam_z(\lambdav))$ with $\Sigmam_z(\lambdav) = \sum_{\ell = 0}^L \lambda_\ell \Phim_\ell$ for some vector of non-negative auxiliary variables $\lambdav \geq \zerov$ and each transmitter has power constraint $\EE[|x_k|^2] \leq p_k$, subject to a total sum-power constraint
\begin{equation} \label{eq:dual-mac-constr}
\sum_{k=1}^K p_k \leq \sum_{\ell=0}^L \lambda_\ell \gamma_\ell
\end{equation}
Then, for any $\lambdav \geq \zerov$, the value of the original MISO-BC WSRM problem is upperbounded by the
value of the new MAC WSRM problem
\begin{eqnarray} \label{eq:wsrm-dual}
\mbox{maximize} & & \sum_{k=1}^K W_k \widehat{R}_k \nonumber \\
\mbox{subject to} & & \widehat{\Rm} \in \Cc_{\rm mac}
\end{eqnarray}
where $\Cc_{\rm mac}$ denotes the capacity region of the dual MAC defined above for given parameters
$\lambdav$, $\{\Phim_\ell\}$ and $\{\gamma_\ell\}$. Furthermore, the upperbound provided by the dual MAC is tight.
Letting the value of the problem (\ref{eq:wsrm-dual}), for given $\lambdav$,  be denoted
by $g(\lambdav)$,  then the weighted rate sum of the MISO-BC can be obtained by minimizing $g(\lambdav)$ with
respect to $\lambdav \geq \zerov$.   Hence, the sought solution can be obtained by iterating between one ``outer problem'', for the minimization of $g(\lambdav)$,
and an ``inner problem'', that solves (\ref{eq:wsrm-dual}) for fixed $\lambdav$.

Owing to the polymatroid structure of the capacity region of the Gaussian MAC \cite{Tse-Hanly-TSP07}, the solution
of (\ref{eq:wsrm-dual}) is found at the vertex of the MAC capacity region dominant face corresponding
to the {\em successive decoding} order $K, K-1, \ldots, 1$.  Then,  $g(\lambdav)$ is obtained by solving
\begin{align} \label{eq:wsrm-sinr-dual-max}
& \max_{\pv \geq 0} \;\; \sum_{k=1}^K W_k \log \frac{\left | \Sigmam_z(\lambdav) + \sum_{j=1}^k \hv_j \hv^\herm_j p_j \right |}{\left |
\Sigmam_z(\lambdav) + \sum_{j=1}^{k-1} \hv_j\hv^\herm_j p_j \right |} \nonumber \\
&\mbox{subject to} \;\;\; \Sigmam_z(\lambdav) = \sum_{\ell=0}^L \lambda_\ell \Phim_\ell, \;\;\; \sum_{k=1}^K p_k \leq
\sum_{\ell=0}^L \lambda_\ell \gamma_\ell.
\end{align}
For fixed $\lambdav$, (\ref{eq:wsrm-sinr-dual-max}) can be solved using the Lagrangian duality approach of \cite{Yu-TIT06}, as
done in \cite{Kobayashi-Caire-ICASSP07}.

As far as the the minimization of $g(\lambdav)$ is concerned, this can be obtained
the sub-gradient update of the auxiliary variables in the form $\lambdav(n+1) = \lambdav(n) - \epsilon_n \; \sv(\lambdav(n))$,
where $\lambdav(n)$ denotes the current value of $\lambdav$ at iteration $n$,
$\sv(\lambdav(n))$ is a subgradient  of $g(\lambdav)$ at $\lambdav = \lambdav(n)$, and $\epsilon_n = \epsilon_0 \frac{1 + b}{n + b}$ is the adaptation step, for some suitable parameters  $\epsilon_0, b > 0$. A subgradient for this problem is given by the vector with components \cite{Zhang-Poor-etal-ISIT09}
\[ s_\ell(\lambdav) = \gamma_\ell - \trace \left ( \Sigmam_x(\lambdav) \Phim_\ell \right ), \]
where $\Sigmam_x(\lambdav)$ denotes the transmit covariance matrix of the MISO-BC corresponding to the dual MAC at given $\lambdav$.
The calculation of the subgradient requires to map the dual MAC solution $\{p_k(\lambdav)\}, \{\widehat{\wv}_k(\lambdav)\}$ into the corresponding
solution (powers and steering vectors) of the MISO-BC in order to determine $\Sigmam_x(\lambdav) = \sum_{k=1}^K \vv_k(\lambdav)\vv_k^\herm(\lambdav) q_k(\lambdav)$.
This is obtained by the well-known ``MAC-to-BC'' transformations \cite{Vishwanath-Jindal-Goldsmith-TIT03}.


An alternative approach to the solution of problem (\ref{eq:wsrm}) can be obtained by extending
the ``min-max duality'' approach of \cite{Yu-Lan-TSP07} to the case of general linear constraints and arbitrary rate weights.
Consider the downlink power minimization problem with SINR constraints and general linear constraints:
\begin{eqnarray} \label{eq:min-pow}
{\rm minimize} && P \nonumber \\
\mbox{subject to} &&  \SINR_k^{\rm dl} \geq \eta_k, \;\;\; \forall k \nonumber \\
&& \trace \left( \sum_{k=1}^K \wv_k \wv_k^\herm \right) \leq P, \nonumber \\
&& \trace \left( \sum_{k=1}^K \wv_k \wv_k^\herm \Phim_\ell \right) \leq \gamma_\ell, \;\;\; \forall \ell
\end{eqnarray}
where $\wv_k = \sqrt{q_k} \vv_k$ denotes the unnormalized downlink beamforming vectors, the downlink SINR for user $k$ is given by
\[ \SINR^{\rm dl}_k = \frac{\left |\hv_k^\herm \wv_k \right |^2}{1 + \sum_{j\neq k} \left |\hv_k^\herm \wv_j\right |^2} \]
and $\eta_k$ denotes the SINR target for user $k$.

\vspace{5pt}
\begin{theorem} \label{thm:min-pow-dual}
The downlink beamforming problem (\ref{eq:min-pow}) has the following Lagrangian dual form which is equivalent to a dual uplink problem
with the same SINR constraints and under a worst-case noise condition:
\begin{eqnarray} \label{eq:min-pow-dual-final}
\max_{\lambdav \geq 0} \; \min_{\pv \geq 0, \{\widehat{\wv}_k\}} && \sum_{k=1}^K p_k - \sum_{\ell=1}^L \lambda_\ell \gamma_\ell \nonumber \\
\mbox{subject to} && \frac{p_k |\widehat{\wv}_k^\herm \hv_k|^2}{\widehat{\wv}_k^\herm \Sigmam'_z(\lambdav) \widehat{\wv}_k + \sum_{j\neq k} p_j |\widehat{\wv}_k^\herm \hv_j|^2 } \geq \eta_k, \;\;\; \forall k \nonumber \\
&& \Sigmam'_z(\lambdav) = \Id + \sum_{\ell=1}^L \lambda_\ell \Phim_\ell
\end{eqnarray}
where $\{p_k\}$ and $\{\lambda_\ell\}$ are the dual variables for the SINR constraint and general linear constraint, respectively.
\end{theorem}
\begin{IEEEproof}
See Appendix \ref{appen:minmaxdual}.
\end{IEEEproof}
\vspace{5pt}

Notice that the sum power in the corresponding dual MISO-BC (downlink) is not explicitly constrained.
In contrast, the objective function is modified by a discount factor that includes the ``noise'' variables $\lambda_1,\ldots,\lambda_L$.
At the optimal point, the MISO-BC sum power is given by $P^* = \sum_{k=1}^{K} p^*_k - \sum_{\ell=1}^L \lambda^*_\ell \gamma_\ell$. Thus, $P^*$ is generally smaller than the dual-MAC sum power $\sum_{k=1}^{K} p^*_k$. This suggests that for a given total power budget $P^*$ (fixed), we may need to reduce the downlink transmit power in order to satisfy the linear constraints. The same duality holds if we consider DPC successive encoding in some given order
(say: $\pi = (\pi_1,\ldots,\pi_K)$)  and successive interference cancelation in the dual-MAC reverse order
(say: $\pi_K$ decoded first and $\pi_1$ decoded last).  Since the capacity region of the multiuser MISO downlink channel is obtained as convex hull of the union
of DPC-achievable regions over all possible transmit covariances
satisfying a general convex constraint (see (\ref{eq:cap-region})), and since the SINRs for each rate point of such regions are also achievable in the dual MAC,
we conclude that the capacity region of the downlink subject to the general linear constraints coincides with the capacity region of a virtual dual MAC with worst-case noise covariance, where the covariance matrix is parameterized by $\Sigmam'_z(\lambdav)$ in the specific form given above.

Consider now the downlink WSRM problem (\ref{eq:wsrm}) where $\Sc$ is defined by general linear constraints
(\ref{eq:lin-constr}), including the sum-power constraint $\trace(\Sigmam_x) \leq P$.
Letting again, without loss of generality, the weights be ordered such that $W_1 \geq \cdots \geq W_K$, the resulting min-max dual MAC problem is given by:
\begin{eqnarray} \label{eq:wsrm-minmax-dual}
\min_{\lambdav \geq 0} \; \max_{\pv \geq 0}  && \sum_{k=1}^K W_k \log \frac{\left | \Sigmam'_z(\lambdav) + \sum_{j=1}^k \hv_j \hv^\herm_j p_j \right |}{\left | \Sigmam'_z(\lambdav) + \sum_{j=1}^{k-1} \hv_j\hv^\herm_j p_j \right |} \nonumber \\
\mbox{subject to} && \Sigmam'_z(\lambdav) = \Id + \sum_{\ell=1}^L \lambda_\ell \Phim_\ell, \nonumber \\
&& \sum_{k=1}^K p_k \leq P + \sum_{\ell=1}^L \lambda_\ell \gamma_\ell
\end{eqnarray}
By comparing (\ref{eq:wsrm-minmax-dual}) with (\ref{eq:wsrm-sinr-dual-max})
and recalling that the value $g(\lambdav)$ of (\ref{eq:wsrm-sinr-dual-max}) must be minimized with respect to $\lambdav \geq \zerov$,
we notice that the only difference between the two formulations is the presence of the auxiliary variable $\lambda_0$
in (\ref{eq:wsrm-sinr-dual-max}), related to the sum-power constraint.  However, it is immediate to see that the solution of (\ref{eq:wsrm-sinr-dual-max})
is invariant to a common scaling  of the vector of auxiliary variables $\lambdav$ since this would affect in the same way both the noise covariance and the
signal power constraint. Without loss of generality, letting $\lambda_0 = 1$ in (\ref{eq:wsrm-sinr-dual-max}) yields the same optimization problem as (\ref{eq:wsrm-minmax-dual}).

Therefore, the infeasible start Newton method of \cite{Yu-Lan-TSP07}
can be used as an alternative to the inner (Lagrange duality) -- outer (subgradient)  iterative algorithm advocated in \cite{Zhang-Poor-etal-ISIT09}.
Since this algorithm is only sketched  in \cite{Yu-Lan-TSP07} for the case of per-antenna power constraint,
and several computation steps are left to  the reader, we give the details here for the case of general linear constraints.
First, we define the modified objective function for (\ref{eq:wsrm-minmax-dual})
\begin{eqnarray} \label{eq:mac-obj}
f_t(\pv,\lambdav) & = & \sum_{k = 1}^K \Delta_k \log \left | \Id + \sum_{\ell=1}^L \lambda_\ell \Phim_\ell + \sum_{j=1}^k \hv_j \hv_j^\herm p_j \right | - 
W_1 \log \left | \Id + \sum_{\ell=1}^L \lambda_\ell \Phim_\ell \right | \nonumber \\
& & + \frac{1}{t} \left ( \sum_{k=1}^K \log p_k - \sum_{\ell=1}^L \log \lambda_\ell \right )
\end{eqnarray}
where $t > 0$ is a parameter that controls a ``logarithmic barrier'' term in order to prevent the iterative algorithm to approach the boundaries where some elements in $\pv$ or in $\lambdav$ may become zero or negative and where we define $\Delta_k = W_k - W_{k+1}$ with $W_{K+1} = 0$.
The logarithmic barrier guarantees that the optimal value of the problem can be approached with gap $\frac{K + L}{t}$.
Along the iterations, the value $t$ shall be increased in order to make this gap as small as desired.

The problem is convex with respect to $\lambdav$ and concave with respect to $\pv$,
with Lagrangian function (neglecting the non-negativity constraints and using the modified objective function (\ref{eq:mac-obj})) given by
\begin{equation} \label{eq:lagrange1}
\Lc(\pv,\lambdav, \mu) = f_t(\pv,\lambdav) - \mu \left ( \onev^\transp \pv - P - \gammav^\transp \lambdav \right )
\end{equation}
with $\gammav = (\gamma_1,\ldots,\gamma_L)^\transp$ and $\pv = (p_1,\ldots,p_K)^\transp$. The necessary and sufficient conditions for optimality are given by the KKT conditions \cite{Boyd-Vandenberghe-2004}:
\begin{eqnarray} \label{eq:KKT-residual}
\rv_1 & = & \frac{\partial f_t(\pv,\lambdav)}{\partial \pv} - \mu \onev = 0 \nonumber \\
\rv_2 & = & \frac{\partial f_t(\pv,\lambdav)}{\partial \lambdav} + \mu \gammav = 0 \nonumber \\
r_3    & = &  P  + \gammav^\transp \lambdav - \onev^\transp \pv = 0
\end{eqnarray}
The vector $\rv = (\rv_1^\transp, \rv_2^\transp, r_3)^\transp$ of dimension $K + L + 1$ is the so-called ``residual'' of the KKT equations.
The algorithm finds a direction and a step for updating the variables $(\pv, \lambdav, \mu) \geq 0$  such that, as the number of iterations grows,  the norm of the residual tends to zero. The updating direction is given by  $\dv = - \left ( \nabla \rv \right )^{-1} \rv$, where
$\nabla \rv $ is the {\em KKT matrix}, given by
\begin{equation} \label{eq:kktmatrix}
\nabla \rv = \left [ \begin{array}{ccc}
\frac{\partial \rv_1}{\partial \pv^\transp} & \frac{\partial \rv_1}{\partial \lambdav^\transp} & - \onev \vspace{3mm} \\
\frac{\partial \rv_2}{\partial \pv^\transp} & \frac{\partial \rv_2}{\partial \lambdav^\transp} & \gammav \vspace{3mm} \\
- \onev^\transp & \gammav^\transp & 0 \end{array} \right ]
\end{equation}
Letting for simplicity the vector of variables be denoted by $\xsf = (\pv^\transp, \lambdav^\transp, \mu)^\transp$, the algorithm
takes the following form:
\begin{enumerate}
\item Fix the algorithm parameters $\nu > 1$, and $\delta > 0$. Initialize $\xsf(0)$ to some positive values and let $n = 0$, and $t = 1$.
\item Compute the updating direction $\dv(n)$ at $\xsf(n)$. (see explicit expressions of the derivatives given later on).
\item Update $\xsf(n+1) = \xsf(n) + s \dv(n)$ where $s$ is found by backtracking line search:
initialize $s = 1$ and find $s$ such as, while
\[ \left \|\rv(\xsf(n) + s \dv(n)) \right \| > (1 - \alpha s) \left \|\rv(\xsf(n)) \right \| \]
then $s \leftarrow \beta s$, where $\beta \in (0,1)$ and $\alpha \in (0,1/2)$ are fixed constants.
(typical values are $\alpha = 0.3$ and $\beta = 0.8$).
\item If $\|\rv(\xsf(n+1)) \| \leq \delta$, move to the next step, otherwise set $n \leftarrow n + 1$ and go back to step 2.
\item If $\frac{K+L}{t} \leq \delta$, exit and accept the value of $\xsf(n+1)$ as the final value, otherwise set $t \leftarrow \nu t$ and $n \leftarrow n + 1$ and go back to step 2.
\end{enumerate}
Explicit expressions for the elements of the KKT matrix $\nabla \rv$ can be obtained using matrix calculus
(see for example \cite{Brewer-TCS78} and references therein).
Letting $\Psim_k = \left [\Id +  \sum_{\ell=1}^L \lambda_\ell \Phim_\ell + \sum_{j=1}^k \hv_j \hv_j p_j \right ]^{-1}$, we find:\footnote{Here $\delta_{i,j}$ denotes Kronecker's delta, equal to 1 if $i=j$ and to 0 otherwise.}
\begin{eqnarray*}
\left [ \frac{\partial \rv_1}{\partial \pv^\transp} \right ]_{i,j} & = &  - \sum_{k = \max\{i,j\}}^K \Delta_k  \hv_i^\herm \Psim_k \hv_j \hv_j^\herm \Psim_k \hv_i  - \frac{\delta_{i,j}}{t p_i^2} \\
\left [ \frac{\partial \rv_1}{\partial \lambdav^\transp} \right ]_{i,j} & = & - \sum_{k = i}^K \Delta_k \hv_i^\herm \Psim_k \Phim_j \Psim_k \hv_i = \left [ \frac{\partial \rv_2}{\partial \pv^\transp} \right ]_{j,i} \\
\left [ \frac{\partial \rv_2}{\partial \lambdav^\transp} \right ]_{i,j} & = & - \sum_{k = 1}^K \Delta_k \trace \left ( \Psim_k \Phim_j \Psim_k \Phim_i \right )
+ W_1 \trace \left ( \Psim_0 \Phim_j \Psim_0 \Phim_i  \right ) + \frac{\delta_{i,j}}{t \lambda_i^2}
\end{eqnarray*}
It should be noticed that the above terms are particularly easily computed in the relevant case where the constraint matrices have
rank 1, i.e., for $\Phim_\ell = \cv_\ell\cv_\ell^\herm$, as in the case of interference direction constraints (see example in Section \ref{sec:multicell}).

\begin{figure}
  \centering
  \includegraphics[width=4.5in]{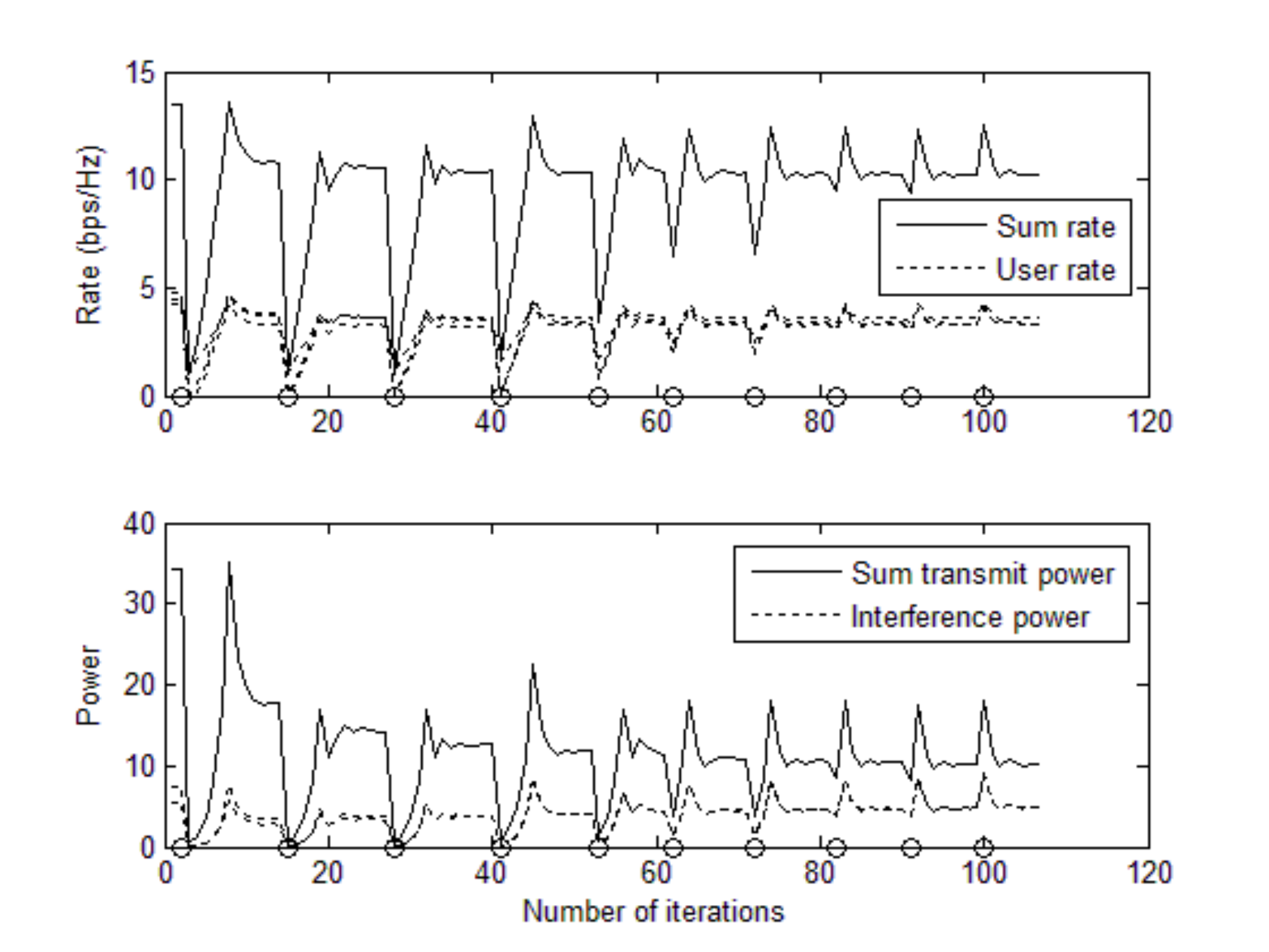}
  \caption{Rate and power convergence behavior of inner-outer iterative algorithm for DPC with $M=4$ and $K=3$ under the sum transmit power and interference constraints with $L=2$ forbidden directions. The dots on the ``x'' axis indicate when the outer subgradient iteration is activated.}
  \label{fig:conv-dpc-inout}
\end{figure}
\begin{figure}
  \centering
  \includegraphics[width=4.5in]{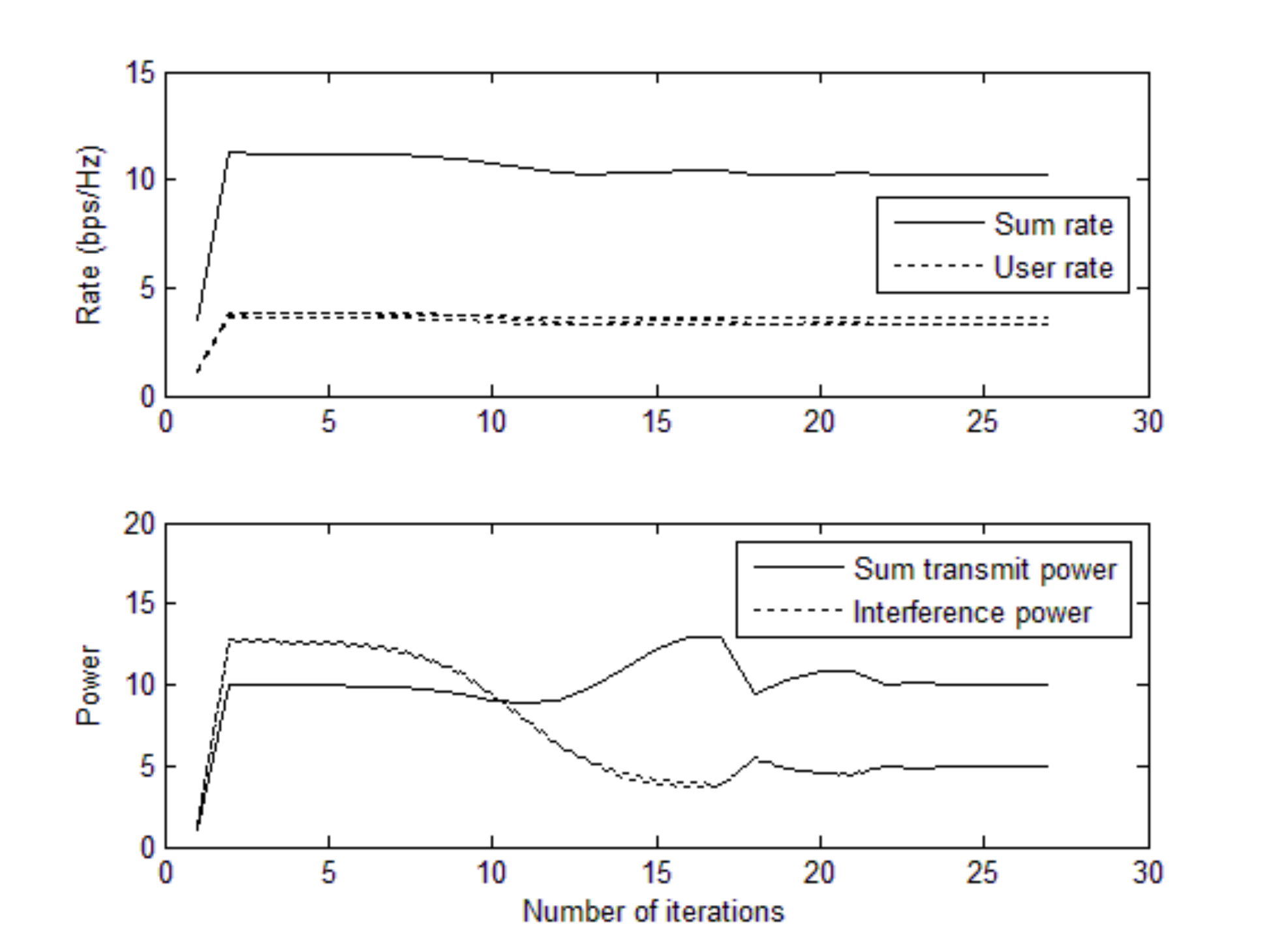}
  \caption{Rate and power convergence behavior of infeasible start Newton algorithm for DPC under the same conditions of Fig.~\ref{fig:conv-dpc-inout}.}
  \label{fig:conv-dpc-newton}
\end{figure}

Figs.~\ref{fig:conv-dpc-inout} and \ref{fig:conv-dpc-newton} show an example of convergence of the inner-outer iterative algorithm
and infeasible start Newton algorithm with the same system parameters with $M = 4$ antennas, $K = 3$ users, unit weights for all users ($W_k = 1$)
and $L = 2$ forbidden interference directions.
In order to allow independent replication of our numerical experiments,
the channel and interference direction vectors  are provided in Table. \ref{tb:example-vectors}.
The sum power constraint is set equal to $10$ and the two interference constraints are set equal to $5$.
The evolution along the algorithm iterations of the objective function (sum rate) and of the sum-power and interference values are shown.
As the figures reveal, both algorithms converge to the same optimal values and satisfy the given sum-power and interference power constraints with equality.
However the workload required to achieve sufficient convergence is different for each algorithm.
The number of $M\times M$ matrix multiplications per iteration
captures to first order the overall algorithms complexity, as the workload of the remaining operations is significantly lower.
The infeasible start Newton algorithm has slightly higher complexity per iteration,
$O(K(K+L^2))$,  than the complexity of the inner-outer iterative algorithm, which is $O(K^2)$.
However, this difference is almost negligible in the relevant case where $K$ is significantly larger than $L$.\footnote{Typically, the number of users per cell $K$ is much larger than
the number of constrained interference direction $L$.}
Furthermore, as seen from Figs.~\ref{fig:conv-dpc-inout} and \ref{fig:conv-dpc-newton}, the Newton algorithm requires a significantly
smaller number of iterations to converge. Therefore, the Newton algorithm has a clear advantage in the case $K \gg L$. For example,
using a MATLAB implementation on a Intel Core 2 Windows XP machine,
for the snapshots of Figs.~\ref{fig:conv-dpc-inout} and \ref{fig:conv-dpc-newton} the infeasible start Newton algorithm
run-time is 58 ms while the inner-outer iterative algorithm run-time is 197 ms (about 3 times slower).
A similar advantage was noticed in a large number of Monte Carlo experiments (not reported here for because of space limitations)
with randomly generated channel vectors.

\begin{table}
  \begin{center}
    \caption{Example of channel and interference direction vectors}
    \begin{tabular}{ |c|c|c||c|c| }
      \hline
      $\hv_1$ & $\hv_2$ & $\hv_3$ & $\cv_1$ &$\cv_2$ \\
      \hline
      $-0.70 + 0.82i$ & $0.20 - 1.10i$ & $0.30 - 0.22i$ & $-0.83 + 0.81i$ & $-0.53 + 0.44i$ \\
      $0.09+ 0.11$ & $-0.70 + 0.90i$ & $-0.50 - 0.65i$ & $0.78 + 0.87i$ & $1.33 - 0.26i$ \\
      $1.15 + 0.04i$ & $0.42 - 0.51i$ & $0.87 - 0.76i$ & $0.45 - 0.45i$ & $0.27 + 0.39i$ \\
      $-0.95 + 0.77i$ & $1.00 - 0.18i$ & $-0.77 - 1.13i$ & $0.78 + 0.55i$ & $-0.83 - 0.87i$ \\
      \hline
    \end{tabular}
    \label{tb:example-vectors}
  \end{center}
\end{table}

\section{Novel WSRM algorithms for ZFBF} \label{sec:zf}

The WSRM problem with ZFBF (\ref{eq:wsrm-zfbf}) can be reformulated in terms of
unnormalized transmit matrices (i.e., {\em including} the transmit powers) as
\begin{eqnarray} \label{eq:wsrm-zfbf1}
\mbox{maximize} &  & \sum_{k=1}^K W_k \log\left (1 +  \hv_k^\herm \Tm_k \hv_k\right ) \nonumber \\
\mbox{subject to} &  & \hv_j^\herm \Tm_k \hv_j = 0 \;\;\;\; \forall j \neq k \nonumber \\
& & \trace \left (\sum_{k=1}^K \Tm_k \Phim_\ell \right ) \leq \gamma_\ell, \;\;\; \forall \ell  \nonumber \\
&  & \Tm_k \succeq 0, \;\;\; {\rm rank}(\Tm_k) = 1, \;\;\; \forall k
\end{eqnarray}
Problem (\ref{eq:wsrm-zfbf1}) is not convex due to the rank-1 constraint.
A convex relaxation of the original problem is obtained by removing the rank-1 constraint.
In \cite{Wiesel-Eldar-Shamai-TSP08} the problem is solved for the equal-weight case and per-antenna constraint
and it is shown that the convex relaxation problem has always a rank-1 solution.
Following in the footsteps, it is easy to show that the same holds for the general case (\ref{eq:wsrm-zfbf1}).
In particular,  letting $\{\Tm_k^*\}$ denote a solution of the convex relaxation
problem with possibly rank$(\Tm_k^*) > 1$ for some $k$, a rank-1 solution $\Tm_k = \tv_k \tv_k^\herm$ achieving the same
optimal value can be determined by finding, independently for each $k$, the vector $\tv_k$ solution of:
\begin{eqnarray} \label{eq:socp}
\mbox{maximize} && \hv_k^\herm \tv_k \nonumber \\
\mbox{subject to} && \hv_k^\herm \tv_k \in \RR_+ \nonumber \\
&& \hv_j^\herm \tv_k = 0, \;\;\; \forall \; j \neq k \nonumber \\
&& \trace \left ( \tv_k \tv_k ^\herm \Phim_\ell \right ) \leq \trace \left (\Tm_k^* \Phim_\ell \right ), \;\;\; \forall \; \ell
\end{eqnarray}
We notice that (\ref{eq:socp}) is a Second-Order Cone Program (SOCP) \cite{Lobo-Vandenberghe-Boyd-Lebret-98}
and can be easily solved by standard tools (e.g., see \cite{YALMIP}). In the special case of per-antenna constraints,
treated in \cite{Wiesel-Eldar-Shamai-TSP08}, (\ref{eq:socp}) reduces to a linear program.

Two main issues arise from the convex relaxation approach:
1) A dramatic dimensionality increase: the relaxed problem deals with $K$ symmetric matrices of dimension
$M \times M$, that is, with $K M(M-1)/2 = O(KM^2)$ variables, in contrast with the $KM$ original variables;
2) Lack of an efficient computational method: in \cite{Wiesel-Eldar-Shamai-TSP08} the relaxed problem for equal weights takes on the form of a ``MAXDET'' \cite{Vandenberghe-Boyd-Wu-SIAM98} for which efficient solvers exist. Unfortunately, for general weights, the problem is not MAXDET and general-purpose
convex optimizers must be used, with consequent increase of the computation burden. In the following we address both issues.
First, we consider a dimensionality reduction of the original problem by eliminating the zero-forcing constraints.
Then,  we propose a  gradient descent algorithm with logarithmic barrier that converged directly to the solution of the dimension-reduced convex relaxation.
Finally, we build on the structure of generalized inverses \cite{Wiesel-Eldar-Shamai-TSP08} and find a low-complexity two-step iterative algorithm where the powers and steering vectors are alternatively updated. The low-complexity algorithm may converge to a local maximum, but if this is combined with a few gradient
descent steps the optimal solution can be approached with very high probability and with a dramatic complexity reduction.

\subsection{Gradient descent algorithm with barrier functions}

We start by reducing the dimensionality of (\ref{eq:wsrm-zfbf1}) by eliminating the ZF constraints.
The condition $\hv_j^\herm \Tm_k \hv_j = 0$ for all $j \neq k$ together with the fact that $\Tm_k \succeq 0$ and $\text{rank}(\Tm_k) = 1$ yields that
\begin{equation} \label{eq:orthocomp}
\Tm_k = \Um_k \av_k \av_k^\herm \Um_k^\herm
\end{equation}
where $\Um_k \in \CC^{M \times (M-K+1)}$ is a unitary basis for the orthogonal complement of the subspace
$\rm{Span}\{\hv_j:j \neq k\}$. Consider the SVD of $\Hm$ in "compact form," i.e., $\Hm = \Um \Sm \Vm^\herm$
with unitary $\Um, \Vm$ of dimensions $M \times K$  and $K \times K$, respectively, and let $\Um^\perp$ be a unitary matrix of
dimension $M \times (M-K)$ such that $[\Um|\Um^\perp]$ is a unitary basis for $\CC^M$.
In particular, $\Um^\perp$ is a basis for the orthogonal complement of $\rm{Span}\{\hv_1 \cdots \hv_K\}$.
The Moore-Penrose pseudoinverse of $\Hm^\herm$ is defined by
\begin{equation} \label{moore-penrose}
\Hm^+ = \Hm(\Hm^\herm \Hm)^{-1} = \Um \Sm^{-1} \Vm^\herm.
\end{equation}
It follows that the $k$-th column of $\Hm^+$ is a linear combination of the columns of $\Um$ and, in addition, it is orthogonal
to all $\hv_j$ for $j \neq k$. Hence, the $k$-th normalized column of $\Hm^+$,  denoted by $\gv_k$ in the following, is a unit-norm vector
in the orthogonal complement  of $\rm{Span}\{\hv_j:j \neq k\}$. Since $\gv_k$ is a linear combination of the columns of $\Um$, then it is also
orthogonal to all columns of $\Um^\perp$. Hence, the desired matrix  $\Um_k$ can be obtained in the form
\[ \Um_k = [\gv_k|\Um^\perp]. \]
Notice that $\hv_k^\herm \Tm_k \hv_k = \hv_k^\herm (\Um_k \av_k \av_k^\herm \Um_k^\herm) \hv_k = | \gv_k^\herm \hv_k |^2 [\Am_k]_{1,1}$,
where we define the rank-1 matrices $\Am_k = \av_k \av_k^\herm$. Letting $d_k = | \gv_k^\herm \hv_k |^2$ and $\widetilde{\Phim}_{\ell,k} = \Um_k^\herm \Phim_{\ell} \Um_k$ for all $k$ and $\ell$, the dimensionality-reduced problem corresponding to (\ref{eq:wsrm-zfbf1}) can be written as
\begin{eqnarray} \label{eq:wsrm-zfbf-nozf}
{\rm maximize}      & & \sum_{k=1}^K W_k \log\left(1+ d_k  [\Am_k]_{1,1} \right) \nonumber \\
\mbox{subject to}  & & \trace \left( \sum_{k=1}^K \Am_k \widetilde{\Phim}_{\ell,k} \right) \leq \gamma_\ell, \; \forall \ell, \nonumber \\
& &  \Am_k \succeq 0, \;\;\; {\rm rank}(\Am_k) = 1, \;\;\; \forall k.
\end{eqnarray}
Again, a convex relaxation of the above problem is obtained by removing the rank-1 constraints.
For the convex relaxation, by including all constraints into logarithmic barrier functions, we obtain the modified objective function
\begin{eqnarray} \label{eq:zfbf-graddes}
f_t(\Am_1, \cdots, \Am_K) & = & \sum_{k=1}^K W_k \log\left(1+d_k [\Am_k]_{1,1} \right) \nonumber \\
& & + \frac{1}{t} \left( \sum_{\ell=1}^L \log \left( \gamma_\ell - \trace \left(\sum_{k=1}^K \Am_k \widetilde{\Phim}_{\ell,k} \right) \right) + \sum_{k=1}^K \log \det \left( \Am_k \right) \right)
\end{eqnarray}
where $t>0$ is the logarithmic barrier control parameter, that guarantees that the optimal value of the problem can be approached with gap $\frac{K + L}{t}$. The problem is concave with respect to $\Am_1, \cdots, \Am_K$ in the domain $\mathbf{dom} \; f_t = \{(\Am_1, \cdots, \Am_K): \trace (\sum_{j=1}^K \Am_j \widetilde{\Phim}_{\ell,j} ) \leq \gamma_\ell, \;\; \Am_k \succeq 0, \;\; \forall \ell, \; k \}$.
We maximize (\ref{eq:zfbf-graddes}) by applying the iterative gradient descent method for given $t$, and increase the parameter $t$ after a sufficient number of iterations. Since $\Am_k$ is a Hermitian matrix with $M-K+1$ real variables on the diagonal and $(M-K)(M-K+1)/2$ complex variables off-diagonal, the problem has a total of $K(M-K+1)^2$ real variables. For $K = M$, this represents a very significant dimensionality reduction with respect to the original convex relaxation of \cite{Wiesel-Eldar-Shamai-TSP08} (from cubic to linear in the number of antennas). However, in applications involving user selection \cite{Yoo-Goldsmith-JSAC06, Dimic-Sidiropoulos-TSP05, Tomasoni-Caire-Ferrari-Bellini-ISIT09} or when the number of antennas is significantly larger than the number of users,
the complexity of the above method is still significant.
In addition, we observed a very slow convergence (see examples later on).
Therefore, we will explore a lower-complexity iterative method in the next section.

For the sake of completeness, we conclude this section by giving explicitly the details of the gradient method.
The partial derivatives of (\ref{eq:zfbf-graddes}) with respect to each element of $\Am_k$ are given by:
\begin{eqnarray} \label{eq:zfbf-gd-subgrad1}
\frac{\partial f_t}{\partial \left[ \Am_k \right]_{\ell,\ell}} = \left\{ \begin{array}{ll}
\frac{W_k d_k}{1+\widetilde{\hv}_k^\herm \Am_k \widetilde{\hv}_k} + \frac{1}{t} \left( -\sum_{\ell=1}^L \frac{[ \widetilde{\Phim}_{\ell,k} ]_{m,m}}{\gamma_\ell - \trace \left(\sum_{j=1}^K \Am_j \widetilde{\Phim}_{\ell,j} \right)} + [\Am_k^{-1}]_{m,m} \right), & m=1 \\
\frac{1}{t} \left( -\sum_{\ell=1}^L \frac{[ \widetilde{\Phim}_{\ell,k} ]_{m,m}}{\gamma_\ell - \trace \left(\sum_{j=1}^K \Am_j \widetilde{\Phim}_{\ell,j} \right)} + [\Am_k^{-1}]_{m,m} \right), & \forall m \neq 1
\end{array} \right.
\end{eqnarray}
\begin{eqnarray} \label{eq:zfbf-gd-subgrad2}
\frac{\partial f_t}{\partial \Re \left([\Am_k]_{m,n}\right)} &=& \frac{1}{t} \left( -\sum_{\ell=1}^L \frac{2 \Re \left( [\widetilde{\Phim}_{\ell,k}]_{m,n} \right)}{\gamma_\ell - \trace \left(\sum_{j=1}^K \Am_j \widetilde{\Phim}_{\ell,j} \right)} + 2\Re\left([\Am_k^{-1}]_{m,n} \right)\right) \nonumber \\
&=& \frac{\partial f_t}{\partial \Re \left([\Am_k]_{n,m}\right)}, \;\;\; \forall m \neq n
\end{eqnarray}
\begin{eqnarray} \label{eq:zfbf-gd-subgrad3}
\frac{\partial f_t}{\partial \Im \left([\Am_k]_{m,n}\right)} &=& \frac{1}{t} \left( -\sum_{\ell=1}^L \frac{2 \Im \left( [\widetilde{\Phim}_{\ell,k}]_{m,n} \right)}{\gamma_\ell - \trace \left(\sum_{j=1}^K \Am_j \widetilde{\Phim}_{\ell,j} \right)} + 2\Im\left([\Am_k^{-1}]_{m,n} \right)\right) \nonumber \\
&=& -\frac{\partial f_t}{\partial \Im \left([\Am_k]_{n,m}\right)}, \;\;\; \forall m \neq n
\end{eqnarray}
The update direction matrix for $\Am_k$ is denoted by $\Dm_k = \nabla_{\Am_k} f_t$, with elements given by
\begin{eqnarray}
[\Dm_k]_{m,n} = \left\{ \begin{array}{ll}
\frac{\partial f_t}{\partial \left[ \Am_k \right]_{m,m}}, & \forall m=n \\
\frac{\partial f_t}{\partial \Re \left([\Am_k]_{m,n}\right)} + j\frac{\partial f_t}{\partial \Im \left([\Am_k]_{m,n}\right)}, & \forall m \neq n
\end{array} \right.
\end{eqnarray}
At the $n$-th iteration of the gradient descent algorithm, the $k$th matrix is updated as $\Am_k(n+1) = \Am_k(n) + s\Dm_k(n)$ where the step size $s$
is determined according to a standard backtracking line search: initialize $s=1$ and update $s \leftarrow \beta s$ while
\[f_t(\Am_1(n+1), \cdots, \Am_K(n+1)) < f_t(\Am_1(n), \cdots, \Am_K(n)) + \alpha s \sum_{k=1}^K \sum_{i \geq j} | [\Dm_k(n)]_{i,j}|^2 \]
or
\[(\Am_1(n+1), \cdots, \Am_K(n+1)) \notin \mathbf{dom} \; f_t,\]
where $\beta \in (0,1)$ and $\alpha \in (0,1/2)$ are fixed constants. For the given control parameter $t$, the matrices $\Am_k (n), \forall k$ are updated until the following stopping criterion is satisfied for the convergence of the objective function in (\ref{eq:zfbf-graddes}):
\[ s \left( \sum_{k=1}^K \sum_{i \geq j} | [\Dm_k(n)]_{i,j}|^2 \right) ^{1/2} < \delta \]
When the stopping criterion is satisfied, $t$ is updated as $t = \nu t$ for $\nu > 1$ and a new gradient descent iteration starts with new $t$.
The solution can be mapped into a rank-1 equivalent solution by letting $\Tm_k^* = \Um_k \Am_k^* \Um_k^\herm$,
where $\{\Am_k^*\}$ denotes the optimal point found by the gradient descent, and then finding the optimal steering vectors $\{\tv_k\}$ via
(\ref{eq:socp}) for each $k$.

Some terms in the logarithmic barrier function may approach the negative infinity as the iterations
proceed. Hence, the algorithm parameters must be designed very conservatively,
allowing a very small step size at each iteration.
For this reason, the gradient descent algorithm converges very slowly.
Fig. \ref{fig:conv-zfbf-graddes} illustrates the convergence behavior of the gradient descent algorithm.
The channel and constraint parameters including the channel and interference direction vectors are the same as in Figs. \ref{fig:conv-dpc-inout} and \ref{fig:conv-dpc-newton}. The sum rate converges to the optimal values and the given sum-power and interference power approaches the given constraints with equality,
but convergence is very slow.

\subsection{Two-step power and steering vector update algorithm} \label{ziocanaglia}

We consider a new algorithm that builds on the structure of generalized inverses, and updates directly the steering vectors rather than working with the convex relaxation problem. In this way, the dimensionality of the problem is not expanded, but the possibility of converging to a local maximum exists.
This problem will be addressed at the end of this section.
In general, the zero-forcing constraint implies that the matrix $\Tm = [\tv_1, \ldots, \tv_K]$ of unnormalized steering vectors must be a
right generalized inverse \cite{Wiesel-Eldar-Shamai-TSP08} of the matrix $\Hm^\herm$, i.e., it can be expressed in the form
\begin{eqnarray} \label{T-form}
\Tm & = & [\gv_1 a_1, \ldots, \gv_K a_K] + \Um^\perp [\bv_1, \ldots, \bv_K] \nonumber \\
& = & \Gm + \Um^\perp \Bm
\end{eqnarray}
where $\{\gv_k\}$ are the normalized columns of the Moore-Penrose pseudo-inverse (\ref{moore-penrose})
and $\Um^\perp$ is a unitary basis of the orthogonal complement of $\rm{Span}\{\hv_1 \cdots \hv_K\}$, as defined before, where $\av = (a_1,\ldots, a_K)^\transp$ are scalar coefficients, and $\Bm$ is a matrix of size $(M - K) \times K$.
We seek to directly optimize the coefficients $\av$  and $\Bm$ by iterating two steps:
1) for fixed (normalized) steering vectors, optimize the power allocation;
2) for fixed power ratios (relative powers) on the pseudo-inverse $\{\gv_k\}$ directions, maximize a common scaling factor by optimizing the steering vectors.

{\bf Step 1.} Initialize the steering vectors by letting $\tv_k = \gv_k$, corresponding to $\av = \onev$ and $\Bm = \zerov$. The ZFBF power allocation problem for fixed (not necessarily unit-norm) steering vectors is given by:
\begin{eqnarray} \label{fixed-steering}
\mbox{maximize} & & \sum_{k=1}^K W_k \log(1 + |\hv_k^\herm \tv_k|^2 q_k) \nonumber\\
\mbox{subject to:} & & \sum_{k=1}^K q_k \tv_k^\herm \Phim_\ell \tv_k \leq \gamma_\ell, \;\;\; \forall \ell \nonumber \\
& & \qv \geq \zerov
\end{eqnarray}
Defining the $L \times K$ matrix $\Cm$ with $(\ell,k)$ element $[\Cm]_{\ell,k} = \frac{1}{\gamma_\ell} \tv_k^\herm \Phim_\ell \tv_k$, the constraint can be compactly written as $\Cm \qv \leq \onev$. The Lagrangian for (\ref{fixed-steering}) is
\begin{equation} \label{Lagrangian}
\Lc(\qv, \lambdav) = \sum_{k=1}^K W_k \log(1 + |\hv_k^\herm \tv_k|^2 q_k) - \lambdav^\transp ( \Cm \qv - \onev )
\end{equation}
where $\lambdav \geq 0$ is a vector of dual variables. The KKT conditions for $q_k$ yield the waterfilling-like solution
\begin{equation} \label{waterfilling-like}
q_k(\lambdav) = \left [ \frac{W_k}{\lambdav^\transp \cv_k} - \frac{1}{|\hv_k^\herm \tv_k|^2} \right ]_+
\end{equation}
where $\cv_k$ is the $k$-th column of $\Cm$. Using this into $\Lc(\qv,\lambdav)$, we can solve the dual problem by minimizing $\Lc(\qv(\lambdav),\lambdav)$ with respect to $\lambdav \geq 0$. It is immediate to check that, for any $\lambdav' \geq 0$,
\begin{eqnarray} \label{subgradient-zf}
\Lc(\qv(\lambdav'),\lambdav') & \geq & \Lc(\qv(\lambdav),\lambdav') \nonumber \\
& = &  \Lc(\qv(\lambdav),\lambdav) + (\onev - \Cm \qv(\lambdav))^\transp ( \lambdav' - \lambdav)
\end{eqnarray}
Therefore,  $\sv(\lambdav) = (\onev - \Cm \qv(\lambdav))$ is a subgradient for $\Lc(\qv(\lambdav),\lambdav)$. It follows that the dual problem can be solved by a simple $L$-dimensional subgradient iteration.

{\bf Step 2.} Let $\qv$ denote the output of Step 1 for fixed steering vectors $\{\tv_k\}$.
In this step we fix $\av$ with components $a_k = \sqrt{q_k} \gv_k^\herm \tv_k$,
and search for the steering vectors that maximize a common power scaling factor $\eta$.
Using (\ref{T-form}) we obtain the optimization problem
\begin{eqnarray} \label{max-mu}
\mbox{maximize} & & \eta \nonumber\\
\mbox{subject to:} & & \frac{\eta^2}{\gamma_\ell} \trace \left ( \Tm \Tm^\herm \Phim_\ell \right ) \leq 1 \;\;\; \forall \; \ell
\end{eqnarray}
with solution readily given by
\[ \eta = \frac{1}{\max_{\ell=1,\ldots,L} \sqrt{ \frac{1}{\gamma_\ell} \trace \left ( \Tm \Tm^\herm \Phim_\ell \right )}} \]
where $\Tm$ is calculated as in (\ref{T-form}), for the fixed coefficients $\av$ and for $\Bm$ solution of
\begin{eqnarray} \label{min-t}
\mbox{minimize}_{\Bm, u} & & u \nonumber\\
\mbox{subject to:} & &  \sqrt{ \frac{1}{\gamma_\ell} \trace \left ( \Tm \Tm^\herm \Phim_\ell \right )} \leq u \;\;\; \forall \; \ell
\end{eqnarray}
It is recognized that (\ref{min-t}) is a SOCP with respect to the variables $u$ and $\Bm$, and can be solved by standard efficient tools (e.g., see \cite{YALMIP}).

\begin{figure}
  \centering
  \includegraphics[width=4.5in]{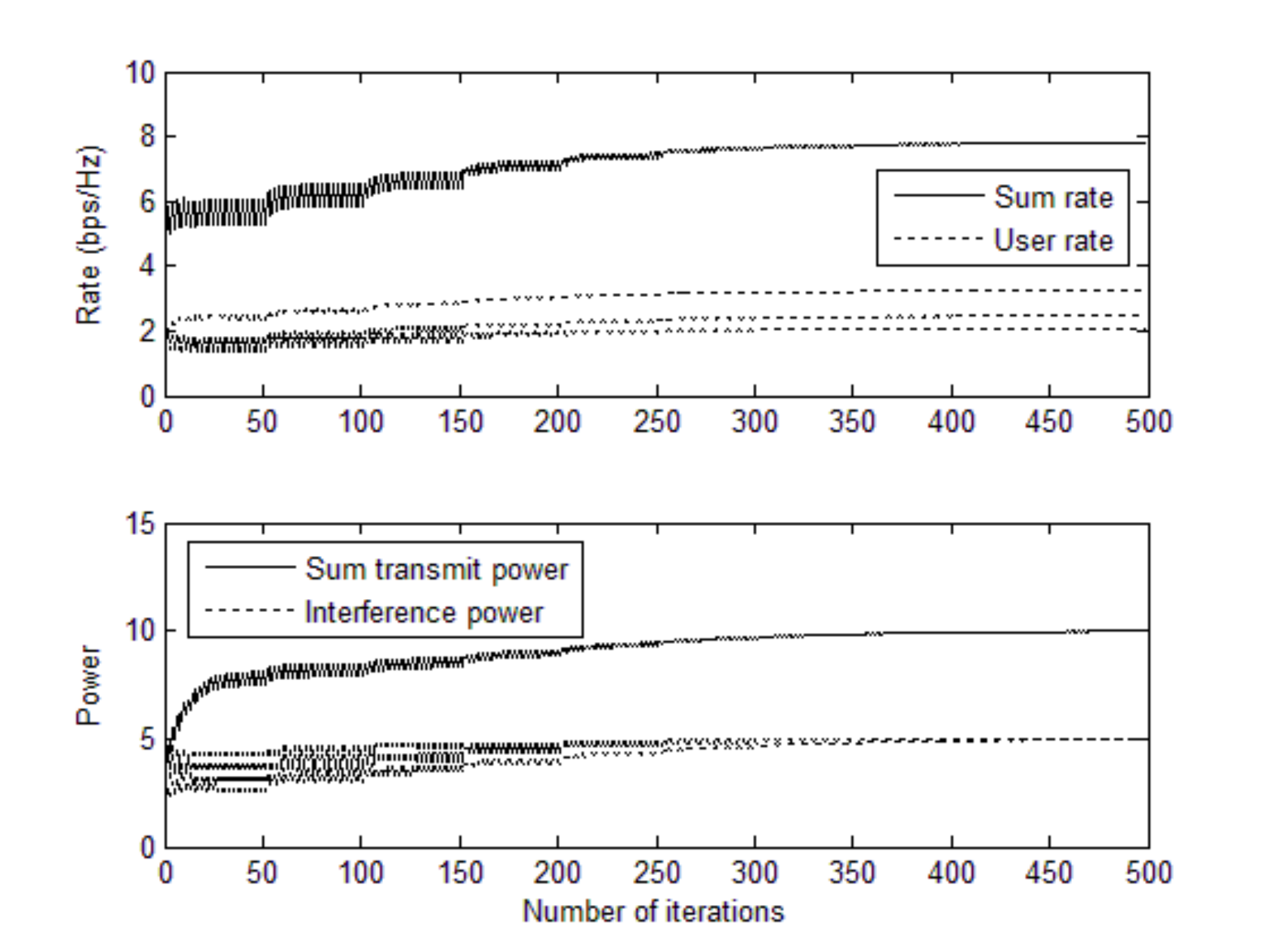}
  \caption{Rate and power convergence behavior of gradient descent algorithm for ZFBF under the same conditions of Fig.~\ref{fig:conv-dpc-inout}.}
  \label{fig:conv-zfbf-graddes}
\end{figure}

\begin{figure}
  \centering
  \includegraphics[width=4.5in]{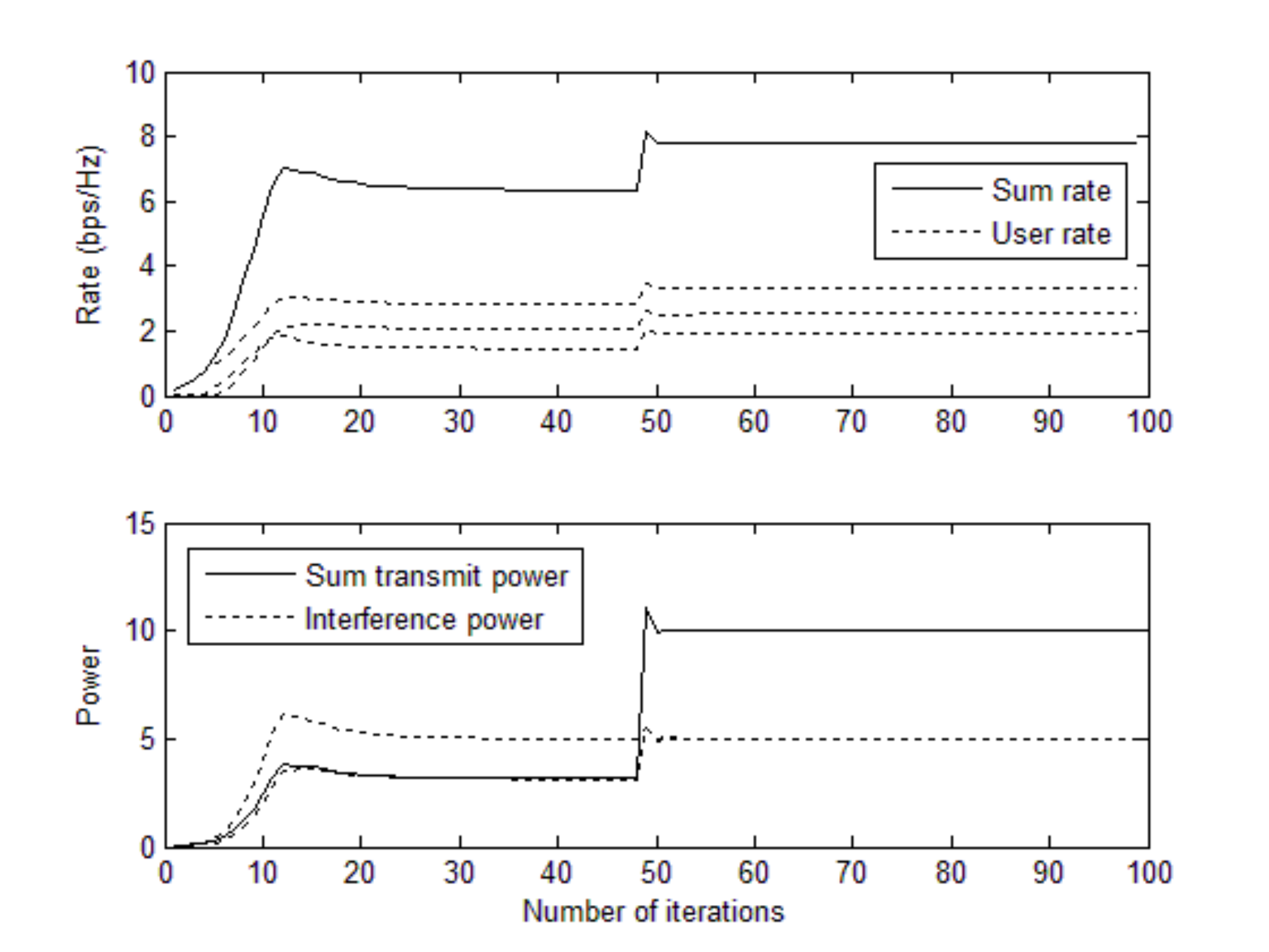}
  \caption{Rate and power convergence behavior of two-step algorithm for ZFBF under the same conditions of Fig.~\ref{fig:conv-dpc-inout}.}
  \label{fig:conv-zfbf-2step}
\end{figure}

The output of Step 2 is a new set of steering vectors in the form $\tv_k = \eta [\gv_k a_k + \Um^\perp \bv_k]$.
These can be used as new fixed steering vectors for Step 1,
and the iterative algorithm can go on.  Notice that, with the initialization $\tv_k = \gv_k$ for all $k$, at the first round of Step 1 the algorithm
obtains the optimal weighted rate sum achievable by the pseudo-inverse steering vectors.
Hence, we are guaranteed to find a generalized inverse that performs at least as well (and usually improves upon)
the pseudo-inverse, already after one iteration.
Fig. \ref{fig:conv-zfbf-2step} shows an example for the two-step algorithm under the same conditions of Fig. \ref{fig:conv-zfbf-graddes}.
In this case, the objective function (sum rate) and the sum-power and interference power converge to the same optimal
values as in Fig. \ref{fig:conv-zfbf-graddes}, but the convergence of the two-step algorithm is significantly faster.

\begin{figure}
  \centering
  \includegraphics[width=5in]{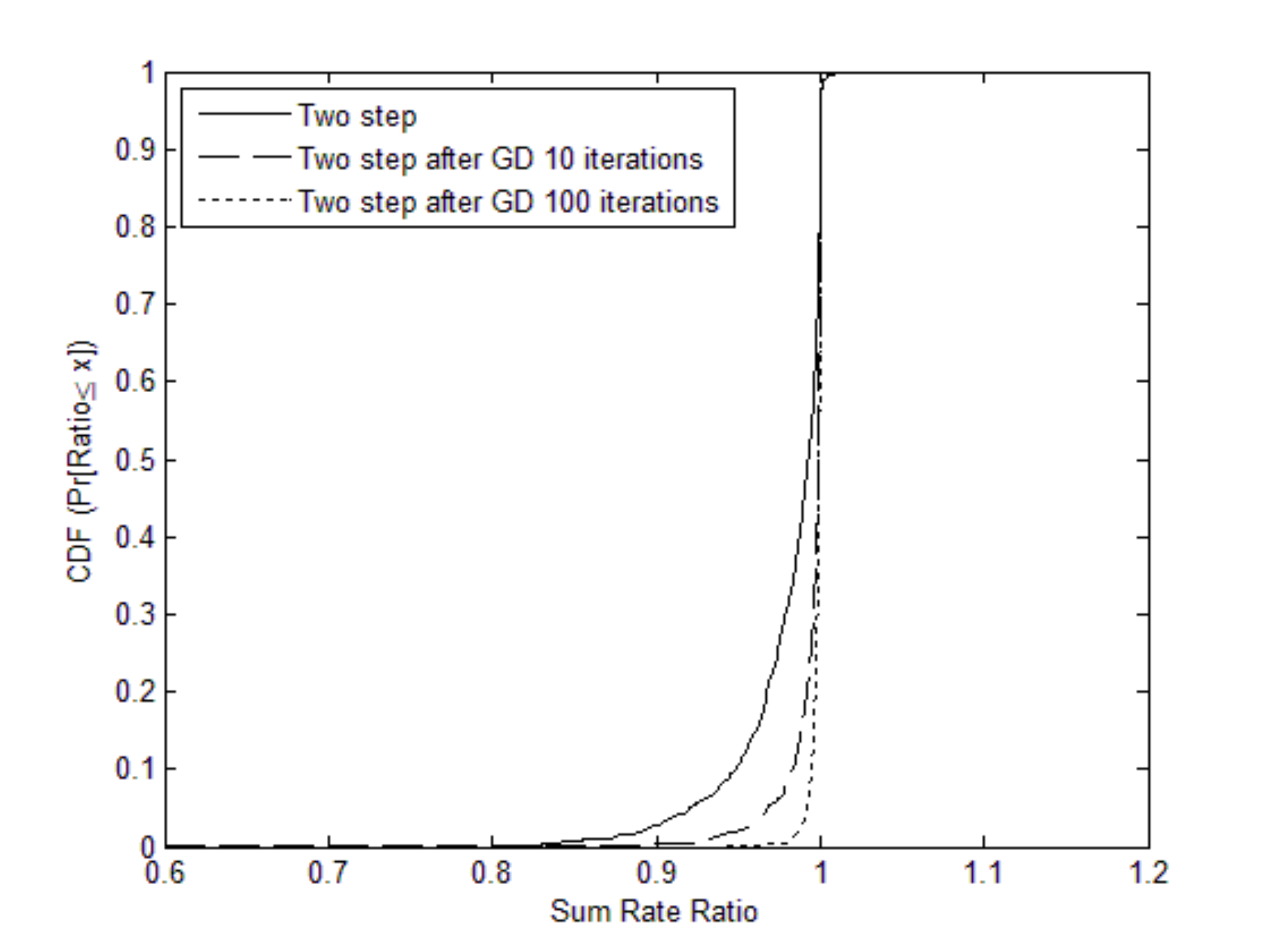}
  \caption{Cumulative distribution of the sum rate of two-step algorithm normalized by the optimal sum rate}
  \label{fig:cdf}
\end{figure}

In general the two-step algorithm may converge to a local maximum since the problem at hand is non-convex.
We investigated this effect by randomly generating a large number of channel matrices with i.i.d. elements $\sim \Cc\Nc(0,1)$ and, for given linear constraints,
we calculated the optimal rate sum obtained using the gradient algorithm and the value achieved by the two-step algorithm.
We assumed $M = 4$ antennas, $K = 3$ users, unit weights for all users,
sum-power constraint  equal to 10 and two interference constraints with random directions and constraint equal to 5.
Fig. \ref{fig:cdf} shows the {\em cumulative distribution function} (CDF) of the ratio between the value of the two-step algorithm
divided by the corresponding optimal value obtained via the gradient algorithm. For example, considering the solid line in Fig. \ref{fig:cdf},
we notice that the two-step algorithm achieves a sum-rate value 5\% less than the optimal with about 10\% probability.
In order to improve the performance of the two-step algorithm, we can use a different initialization point.
Since the gradient method is guaranteed to converge to the optimal point, a sensible choice consists of
performing a limited number of (costly) gradient iterations, and then switching to the (computationally efficient) two-step algorithm.
This approach is meaningful if the feasible point in the convex relaxation problem obtained by the gradient descent algorithm after an arbitrary number of iterations can be mapped into a feasible point for the two-step algorithm, without decreasing the value of the objective function.
This is guaranteed by the following result:

\vspace{5pt}
\begin{theorem} \label{thm:feasible-rank1}
Any feasible set of matrices $\{\widetilde{\Am}_k\}$ of the convex relaxation of problem (\ref{eq:wsrm-zfbf-nozf})
(not necessarily of rank-1) can be mapped into a set of feasible zero-forcing steering vectors $\{\widetilde{\tv}_k\}$ without decreasing the value of the weighted rate sum objective function. This is obtained by solving for each $k$ a SOCP given by (\ref{eq:socp}), with $\Tm_k^*$ replaced by  $\widetilde{\Tm}_k = \Um_k \widetilde{\Am}_k \Um_k^\herm$.
\end{theorem}
\begin{IEEEproof}
See Appendix \ref{appen:feasi-rank1}.
\end{IEEEproof}
\vspace{5pt}

By Theorem \ref{thm:feasible-rank1}, a feasible point obtained after $N$ iterations of the gradient descent algorithm
can be mapped into the initial (feasible) point for the two-step algorithm.
For a sufficiently large number of gradient iterations $N$, the obtained initial point are ``closer'' to the optimal point,
reducing the probability that the two-step algorithm gets trapped into a local maximum.
Fig. \ref{fig:cdf} shows the CDF of the ratio (as defined before) when $N=10$ and $N=100$. We observe
that even with a small number (e.g., 10) gradient iterations, the probability that the two-step algorithm achieves a value very close
to the maximum improves significantly.
This, of course, comes at an enormous saving in complexity with respect to using the gradient method till convergence.

\section{Interference coordination in a multi-cell scenario} \label{sec:multicell}

In cellular wireless communications, the average received signal power is a polynomially decreasing function of the distance between transmitter and receiver.
Users close to the edge of their cell experience relatively weak desired received signal power and strong ICI power and therefore have
very poor SINR. In conventional cellular design \cite{Rappaport-02, Goldsmith-05},
the system capacity is essentially determined by the worst-case ``edge'' users and ICI is mitigated by some fixed allocation
of the downlink transmit power to frequency bands, ranging from the conventional frequency reuse \cite{Rappaport-02, Goldsmith-05} to the
``Fractional Frequency Reuse'' (FFR) schemes advocated in some recent system standardization \cite{Khan-09}.
Such strategies are ``static'' in the sense that they do not exploit the instantaneous knowledge of the interfered users' channel vectors.
The problem of edge users can be alleviated by introducing differentiated
rate services and scheduling. For example, data-oriented high-rate downlink schemes such as EV-DO \cite{Bender-Viterbi-etal-CommMag00}
and HSDPA \cite{Parkvall-Englund-Lundevall-Torsner-CommMag06} consider Proportional Fair Scheduling (PFS)
\cite{Viswanath-Tse-Laroia-TIT02}.

As an application of the transmitter optimization techniques developed before,
in this section we consider a ``partial cell coordination'' approach,
where each base-station is aware of the interfering channel coefficients to users in adjacent cells,
and optimizes its transmitter covariance matrix subject to an {\em interference threshold}
constraint to one or more edge users in adjacent cells.
This approach can be regarded as an intermediate solution between a high complexity fully coordinated
network MIMO approach  \cite{Shamai-Wyner-TIT97, Boccardi-Huang-PIMRC07, Parkvall-Zangi-etal-VTC08F, Caire-Docomo-Allerton08}
and a conventional FFR approach.  We refer to this approach as ``active'' ICI mitigation since it exploits the
instantaneous MIMO channel state information.

In order to fully appreciate the impact of ICI mitigation, the system performance must be evaluated under some suitable
fairness criterion \cite{Bender-Viterbi-etal-CommMag00, Parkvall-Englund-Lundevall-Torsner-CommMag06, Viswanath-Tse-Laroia-TIT02}.
In fact, in a typical setting with $K \geq M$ the cell sum-capacity may be maximized by serving only the users near the cell center,
while allocating zero rate and power to the edge users. This would result in a very misleading result, since the edge users would suffer
from an unacceptably poor quality of service. In our system simulation, we considered PFS and
``Hard-Fairness Scheduling'' (HFS),  where the former aims at maximizing $\sum_{k=1}^K \log \overline{R}_k$ and the latter aims at maximizing $\min_{k=1}^K \overline{R}_k$, where $\overline{R}_k$ denotes the long-term average rate of user $k$. The scheduling algorithms are obtained using the general framework of virtual queues and stochastic optimization presented in \cite{Georgiadis-Neely-Tassiulas-2004} and applied to the MIMO downlink scheduling problem
as done in \cite{ShiraniMehr-Caire-Neely-submit09} where a detailed proof of optimality is also given based on the Lyapunov drift technique.

\subsection{System model} \label{subsec:intfcoordi}

\begin{figure}
  \centering
  \includegraphics[width=5in]{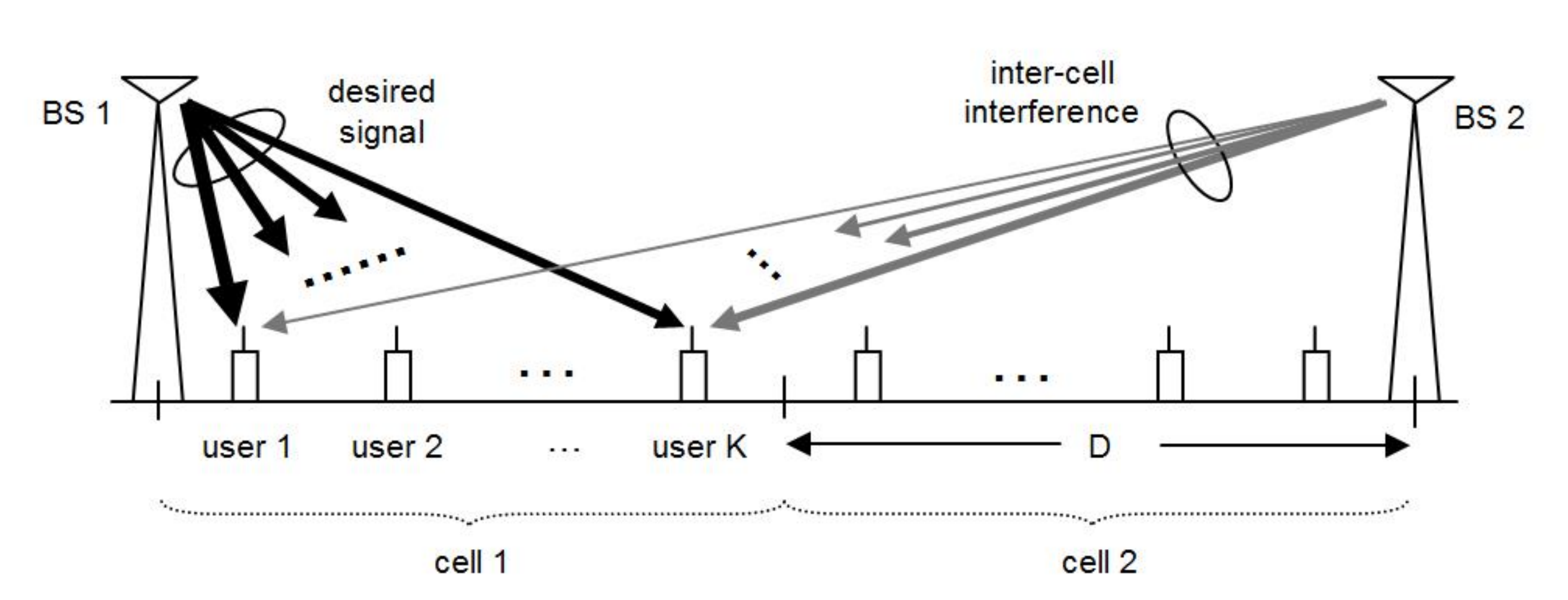}
  \caption{Two-cell multi-user MISO downlink system model.}
  \label{fig:2cellmodel}
\end{figure}

We consider a simple downlink system formed by two mutually interfering cells, as shown in Fig. \ref{fig:2cellmodel}.
Base-stations are placed in position $-D$ and $D$ and serve $K$ users uniformly distributed on the intervals $[-D,0]$ and $[0,D]$, respectively.
In each cell, the users are indexed such that user $k=1$ is the closest to the base-station and user $K$ is at the edge of the cell.
We assume a distance-dependent path gain given by  $G(d) = G_0 / (1 + (d/\delta)^\alpha )$, where $d$ denotes the distance
between the transmitter and receiver, $\alpha$ is the pathloss exponent, $\delta$ is the "3dB" breakpoint distance,
and $G_0$ is a constant  that determines the channel gain at the cell center.
A frequency-flat block-fading channel is assumed, such that at each slot time $t$ the channel vectors are
fixed in time  for the whole slot duration of $T$ channel uses, and then change from slot to slot according to some ergodic process.
The channel vectors have zero-mean i.i.d. Gaussian coefficients (independent Rayleigh fading), both in space (across antennas)
and across the users. The received signal for user $k = 1,\ldots, K$ in cell $n = 1,2$ at slot time $t$ is given by
\begin{eqnarray} \label{eq:mc_reciv}
y_{k,n} (t) &=& \hv_{k,n}^{\herm}(t) \underbrace{\left ( \sum_{j=1}^K \vv_{j,n}(t) u_{j,n} (t) \right )}_{\mbox{from the desired cell}} + \cv_{k,n'}^{\herm}(t)
\underbrace{\left ( \sum_{j=1}^K \vv_{j,n'}(t) u_{j,n'}(t) \right )}_{\mbox{from the interfering cell}} + z_{k,n}(t)
\end{eqnarray}
where $n' = 1$ if $n = 2$ and $n' = 2$ if $n=1$ denotes the neighbor cell index,
$\hv_{k,n}(t)$ and $\cv_{k,n'}(t)$ are the channel vectors from the desired cell to user $k$ antenna and
from the interfering cell to user $k$ antenna, respectively, and where $z_{k,n}(t) \sim \Cc\Nc(0,1)$
denotes a unit-variance AWGN. As before, $\{\vv_{j,n}(t), u_{j,n}(t): j = 1,\ldots, K\}$ and
$\{\vv_{j,n'}(t), u_{j,n'}(t): j = 1,\ldots, K\}$ denote the steering vectors and the coded symbols transmitted by base-station $n$
and $n'$ to their own sets of users, where the dependence on the slot time $t$ is explicitly evidenced.
As described earlier, the vectors $\hv_{k,n}(t)$ have i.i.d. $\Cc\Nc(0, G(d_{k,n}))$ coefficients and the vectors $\cv_{k,n'}(t)$ have i.i.d.
$\Cc\Nc(0, G(s_{k,n'}))$ coefficients, where $d_{k,n}$ denotes the distance between user $k$ in cell $n$ and its desired base-station,
and $s_{k,n'}$ denotes the distance between user $k$ in cell $n$ and the interfering base-station.

Consider cell $n$ (the same operation takes place, independently and symmetrically, in cell $n'$).
At each slot time $t$, given the knowledge of the desired user channels $\{\hv_{j,n}(t): j = 1,\ldots, K\}$ and of the interference
directions to the adjacent users $\{\cv_{j,n}(t): j = 1,\ldots, K\}$, and given the scheduling algorithm that determines
the weights $\{W_{k,n}(t)\}$, the transmitter in cell $n$ optimizes its steering vectors and transmit powers by solving
(\ref{eq:wsrm}) or (\ref{eq:wsrm-zfbf}), depending on whether DPC or ZFBF is considered, subject to the constraints:
\begin{equation} \label{cellular-constraints}
\left \{ \begin{array}{ll}
\trace \left (\Sigmam_x \right ) \leq P,  & \mbox{sum-power} \\
\cv_{K,n}^\herm(t) \Sigmam_x \cv_{K,n}(t)  \leq \epsilon, & \mbox{ICI to the edge user}  \end{array} \right .
\end{equation}
Following the intuition gained by the recent results on the Gaussian interference channel \cite{Etkin-Tse-Wang-TIT08},
the ICI power threshold $\epsilon$ is set equal to the noise level (i.e., equal to 1 in this case),
so that the presence of ICI degrades the edge users' SINR by at most 3 dB.
The user rates resulting from solving (\ref{eq:wsrm}) or (\ref{eq:wsrm-zfbf}) under (\ref{cellular-constraints})
are used to update the scheduling algorithm and to compute the weights to be used in the next slot.
The details of the scheduler equations are omitted for the sake of space limitation and can be found
in \cite{ShiraniMehr-Caire-Neely-submit09}.

By construction, the ICI power for user $K$ is not larger than $\epsilon$.
For all other users $k \neq K$, since $\{\cv_{k,n'}(t): k < K\}$ are independent of
$\cv_{K,n'}(t)$ and of $\{\hv_{j,n'}(t): j =  1,\ldots, K\}$, the average interference power is given by
\[ \EE \left[ \sum_{j=1}^K \left | \cv_{k,n'}^{\herm}(t) \vv_{j,n'}(t) u_{j,n'}(t) \right |^2 \right] = G(s_{k,n'}) P \]
It follows that the noise plus interference power at user $k$ receiver in cell $n$ is given by
\begin{equation} \label{noise-plus-interference}
N_{k,n} = \left \{ \begin{array}{ll}
1 + \epsilon & \mbox{for} \;\; k = K \\
1 + G(s_{k,n'}) P & \mbox{for} \;\; k \neq K \end{array} \right .
\end{equation}
These different ``equivalent noise'' variances can be incorporated as factors in the channel
vectors,  in order  to solve the WSRM problem in the same form as given in the previous sections (details are omitted for brevity).
In our simulations for ZFBF, we use the user selection algorithm that was proposed, under the standard sum-power constraint,
in \cite{Dimic-Sidiropoulos-TSP05} and extend it to the case of non-equal
weights and ICI constraints considered here (details are omitted for the sake of brevity).

\subsection{Simulation results} \label{sec:sim}

In the simulations of this section the cell radius is set to $D = 1$ km and the other system parameters follow
the Mobile WiMAX performance evaluation specification \cite{wimax-eval06}.
Under PFS and HFS, the proposed interference coordination scheme is compared with an FFR interference mitigation scheme
where the total system bandwidth is split into two equal subbands, and the base-stations' transmit power is allocated over the subbands
such that cell 1 uses power  $2P\rho$ in the first subband and $2P(1 - \rho)$ in the second subband for $0 \leq \rho \leq 1$, and cell 2 uses the reverse allocation.
The total base-station power per subband is equal to $P$. However, with this arrangement, the edge users in cell $n$ (where $n=1,2$) can be
served on the higher-power subband, and are interfered by the lower-power subband of the other cell. In the extreme case of  $\rho = 0$, the FFR scheme reduces to classical reuse-2
and for $\rho = 1/2$ we have a  reuse-1 system.
With FFR, we run conventional DPC and ZFBF WSRM problems combined with the scheduling algorithms mentioned
above, subject only to the sum-power constraints on each subband, with the knowledge of the average ICI power
from the adjacent cell,  but no instantaneous knowledge of the interference channels $\{\cv_{k,n}(t), \cv_{k,n'}(t)\}$.
The details of MIMO downlink scheduling in a multi-cell scenario with FFR can be found in \cite{Caire-Docomo-Allerton08}.

\begin{figure}
  \centering
  \includegraphics[width=5in]{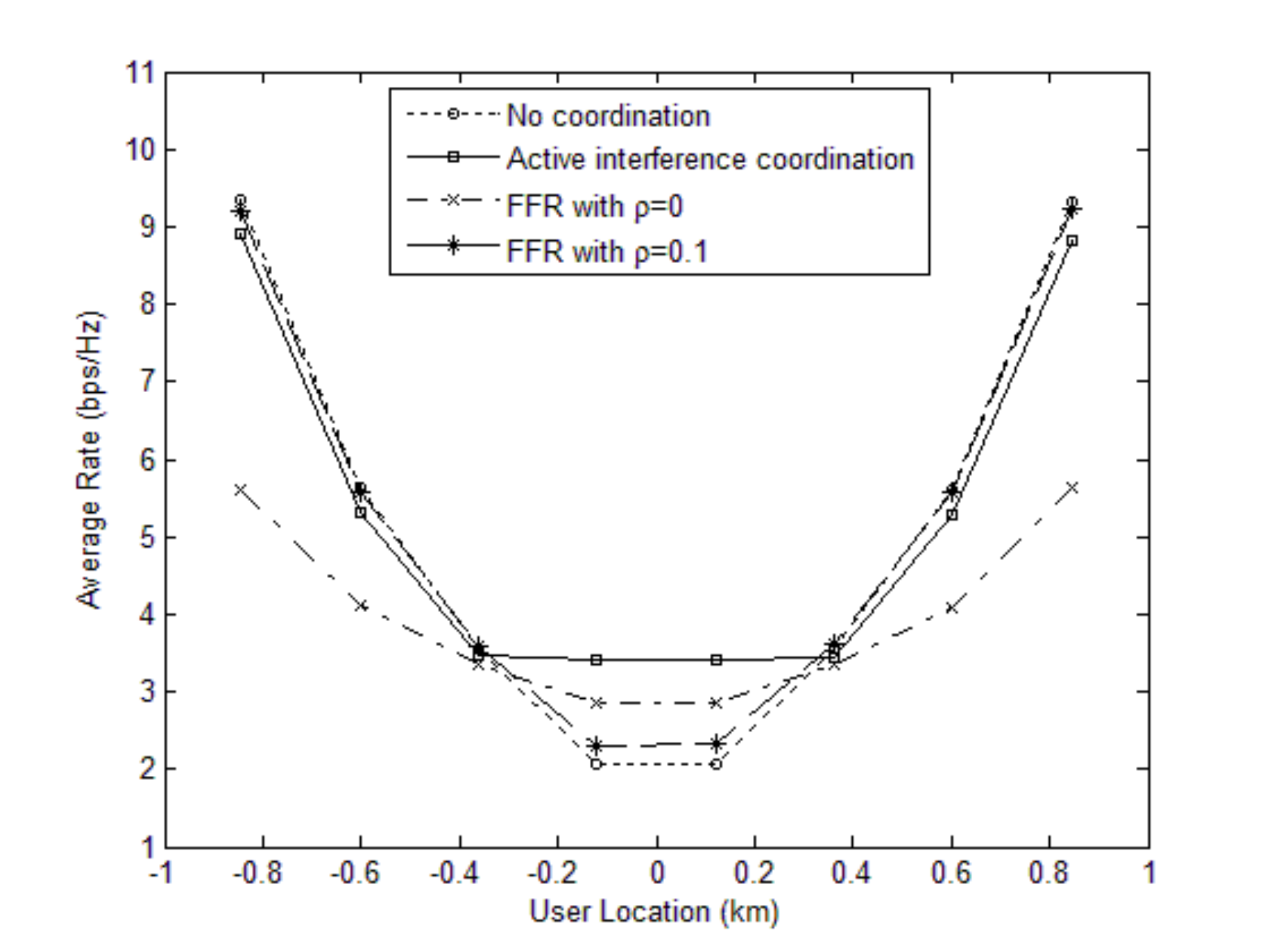}
  \caption{User rate of proportional fairness scheduling for DPC with $M=4$ and $K=4$.}
  \label{fig:pfdpc}
\end{figure}

We considered $M=4$ transmit antennas per base-station, and $K=4$ users per cell.
In the case of active interference mitigation, the ICI constraint (\ref{cellular-constraints}) is imposed, whereby the ICI threshold $\epsilon$ is set equal to 1, i.e., equal to the noise power, as described before.
Based on the typical settings of \cite{wimax-eval06}, the path gain parameters are given as $\alpha=3.504$,
$\delta=0.036$ km, and $G_0=-91.64$ dB and the transmit power normalized by the noise
power is given as $P=154$ dB. Fig. \ref{fig:pfdpc} shows the long-term average user rates in each of the two cells as a function
of the user location for DPC under PFS. For the system where the precoder optimization is subject only to the sum-power constraint, and with frequency reuse 1, the edge user in each cell (i.e., user $K$, located close to $0$) has very small average rate with respect to the center users.
The edge users' rate is significantly improved by the proposed interference coordination, with a negligible
decrease of the center user rates.  Also FFR is able to increase the edge user rates with respect to the no-coordination reuse 1 case,
but the improvement is less significant and comes at the expenses of a rate degradation for the center users.
For example, for  $\rho=0$ (frequency reuse 2), the edge user rates are increased with respect to the $\rho=1$ (frequency reuse 1) case, but the center user rates are reduced by almost a factor of 2.
As a representative example of the FFR performance, we show in the figure the case of $\rho=0.1$.

\begin{figure}
  \centering
  \includegraphics[width=5in]{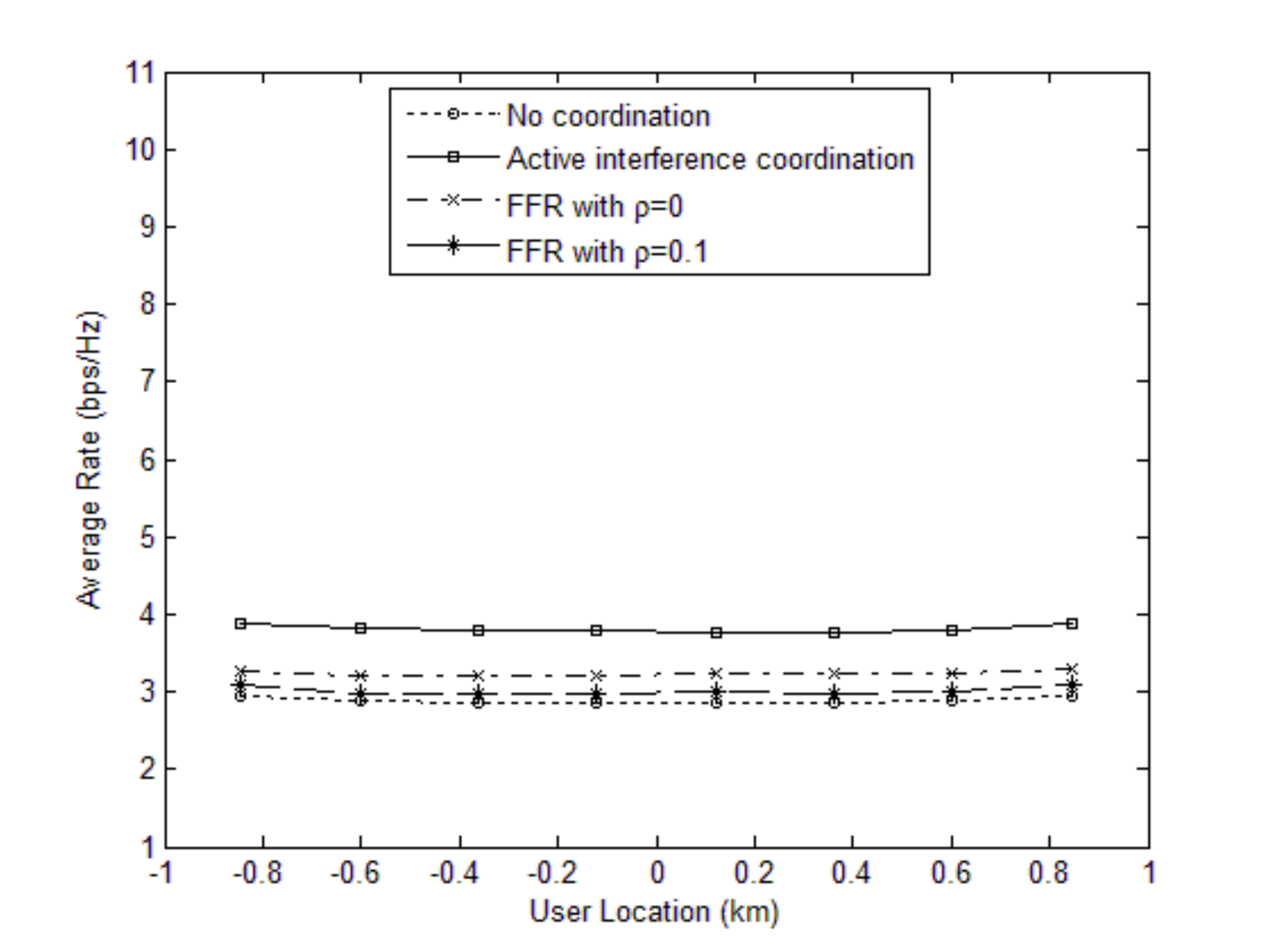}
  \caption{User rate of hard fairness scheduling for DPC with $M=4$ and $K=4$.}
  \label{fig:hfdpc}
\end{figure}

Fig. \ref{fig:hfdpc} shows analogous results for DPC under HFS.
This scheduling scheme achieves indeed the equal long-term average rate point of the system (all users have the same long-term average rate).
Therefore, the bottleneck effect of edge users is even more evident than with PFS.
In this case, the proposed interference coordination scheme achieves the best user rates
and the reuse-2 achieves the best rates over all values of $\rho$ for the FFR case.

\begin{figure}
  \centering
  \includegraphics[width=5in]{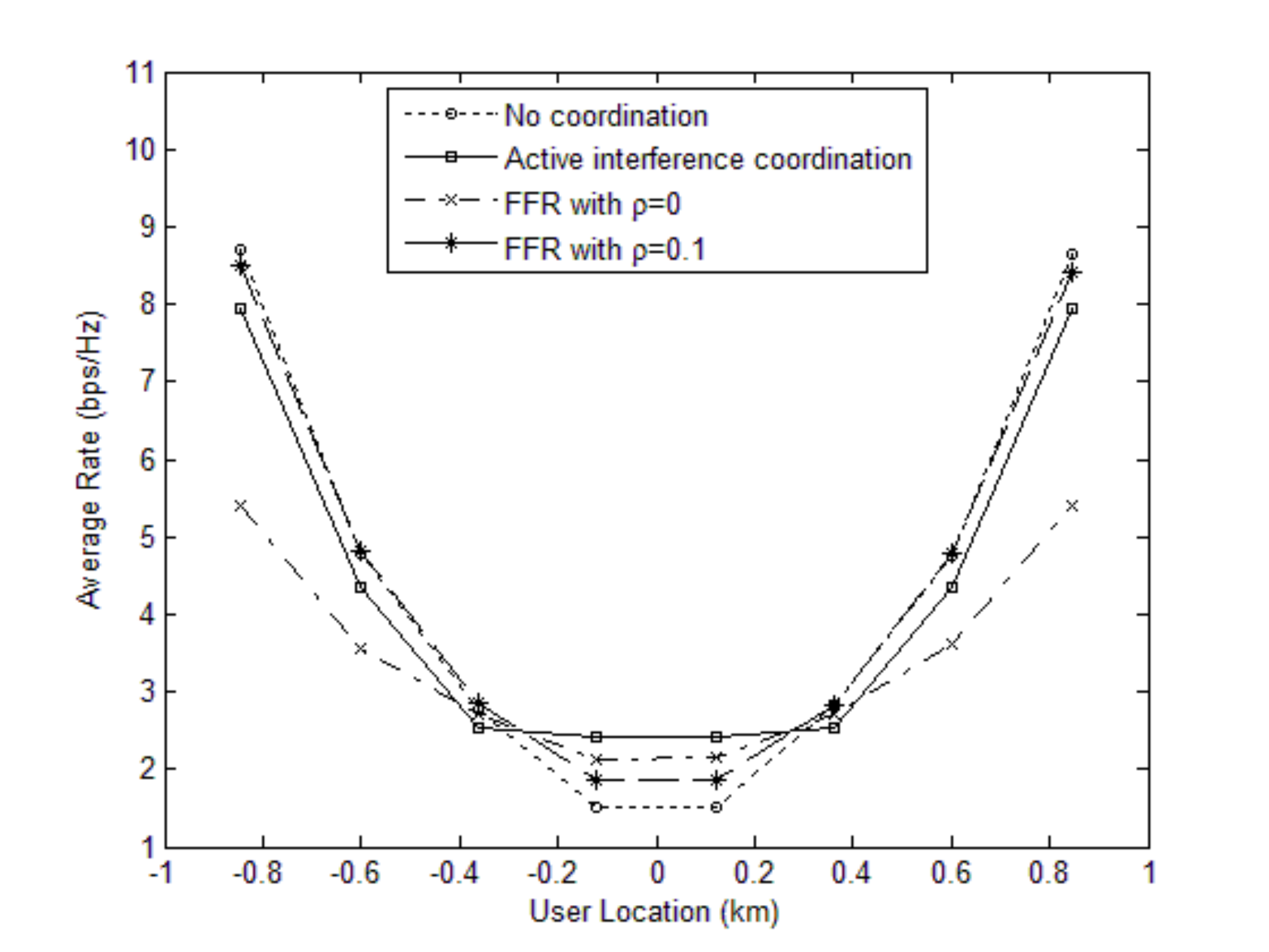}
  \caption{User rate of proportional fairness scheduling for ZFBF with $M=4$ and $K=4$.}
  \label{fig:pfzfbf}
\end{figure}

\begin{figure}
  \centering
  \includegraphics[width=5in]{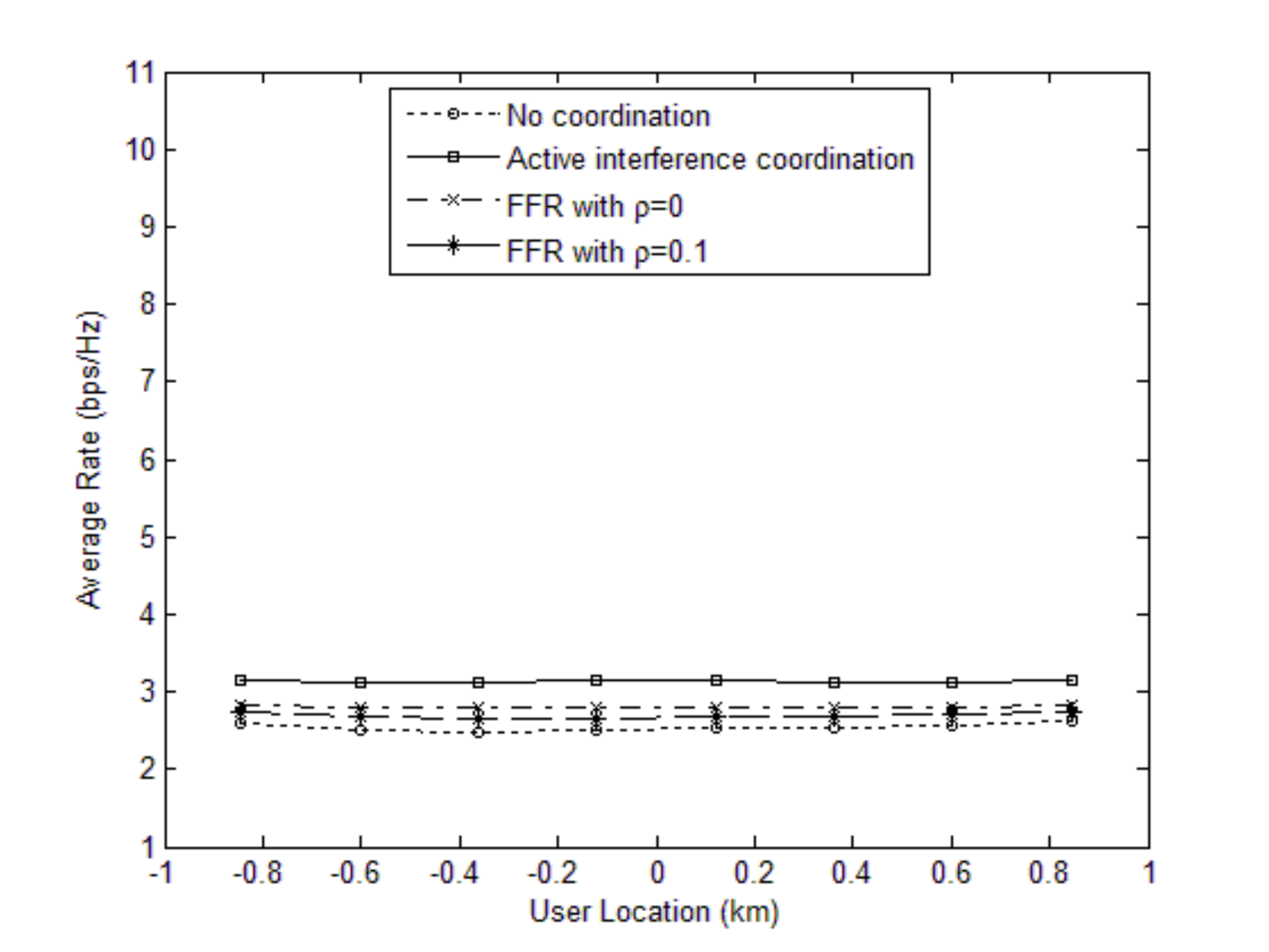}
  \caption{User rate of hard fairness scheduling for ZFBF with $M=4$ and $K=4$.}
  \label{fig:hfzfbf}
\end{figure}

Figs. \ref{fig:pfzfbf}  and \ref{fig:hfzfbf} show the corresponding results for the case of ZFBF under PFS and HFS,
respectively. Observation of these figures reveals trends very similar to those exhibited by DPC precoders in Figs. \ref{fig:pfdpc} and \ref{fig:hfdpc}.

\section{Concluding remarks} \label{sec:conc}

We considered the transmitter optimization problem in a MISO broadcast channel subject to general linear constraints,
under both the optimal DPC and the simpler linear ZFBF precoding schemes.
In this work, we showed the equivalence of ``SINR-duality'' \cite{Zhang-Poor-etal-ISIT09}
and ``min-max duality'' \cite{Yu-Lan-TSP07} approaches, we provided the details of an infeasible start Newton iteration algorithm
and we showed by extensive simulation that it converges significantly faster than the inner-outer iterative algorithm advocated in \cite{Zhang-Poor-etal-ISIT09}.
For the case of ZFBF, the convex relaxation approach of \cite{Wiesel-Eldar-Shamai-TSP08} was extended to the case of general linear constraints and arbitrary
rate weights. We proposed a novel gradient descent algorithm with logarithmic barrier and we addressed the problem of dimensionality
reduction for the convex relaxation problem.  Also, we proposed a novel two-step iterative algorithm that updates directly the steering vectors of the ZF precoder,
building on the form of generalized inverses.  Finally, we  showed an application of these algorithms in a multi-cell ICI mitigation scheme,
where each cell optimizes its transmit covariance matrix by taking into account an interference constraint on the edge users of the adjacent cell.

After the submission of this manuscript, we became aware of two interesting related works that are worth mentioning.
In \cite{zhang1}, linear Zero-Forcing under general linear constraints is addressed for the general case of multi-antenna receivers
(the full MIMO-BC) case. The approach of \cite{zhang1}, particularized to the MISO case, yields a different algorithm that also avoids the dimensionality explosion of
convex relaxation. An accurate complexity comparison between the algorithm of \cite{zhang1} and the one proposed in Section \ref{ziocanaglia}
would be of some interest. In \cite{zhang2}, the $K$-user Gaussian MISO interference channel subject to Gaussian coding, linear beamforming and
treating interference as noise is studied, and a new parameterization of the Pareto-optimal boundary of the beamforming achievable region is
obtained. In this formulation, every transmitter maximizes the rate to its own desired user subject to interference constraints to the other users.
This approach is clearly related to our active interference coordination example. It is expected that by extending the
approach of  \cite{zhang2} to the case of broadcast with interference (where each base station serves many users, as in our case)
a systematic way to set the interference threshold can be found, beyond the sensible ``rule of thumb'' that we used in our example.

\appendices

\section{Proof of Theorem \ref{thm:min-pow-dual}} \label{appen:minmaxdual}

The proof follows closely in the footsteps of \cite{Yu-Lan-TSP07}, generalizing the per-antenna constraint
to arbitrary linear constraints. The Lagrangian function of the power optimization problem (\ref{eq:min-pow}) is given by
\begin{eqnarray} \label{eq:lagrange-minpow}
\Lc(P,\{\wv_k\},\pv,\lambdav) & = & P + \lambda_0 \left [ \trace \left( \sum_{k=1}^K \wv_k \wv_k^\herm \right) - P \right ]
+ \sum_{\ell=1}^L \lambda_\ell \left [ \trace \left( \sum_{k=1}^K \wv_k \wv_k^\herm \Phim_\ell \right) - \gamma_\ell \right ] \nonumber \\
&& - \sum_{k=1}^K p_k \left [ \frac{|\hv_k^\herm \wv_k|^2}{\eta_k} - \sum_{j\neq k} |\hv_k^\herm \wv_j|^2 - 1 \right]
\end{eqnarray}
where $\pv=(p_1, \cdots, p_K)$, $\lambdav=(\lambda_1, \cdots, \lambda_L)$ are the dual variables for the SINR constraints and general linear constraints, respectively. We rewrite (\ref{eq:lagrange-minpow}) as
\begin{eqnarray} \label{eq:lagrange-min-pow1}
\Lc(P,\{\wv_k\},\pv,\lambdav) & = & \sum_{k=1}^K p_k - \sum_{\ell=1}^L \lambda_\ell \gamma_\ell + P ( 1-\lambda_0 ) \nonumber \\
&&  + \sum_{k=1}^K \wv_k^\herm \left( \lambda_0 \Id + \sum_{\ell=1}^L \lambda_\ell \Phim_\ell + \sum_{j \neq k} p_j \hv_j\hv_j^\herm - \frac{p_k}{\eta_k} \hv_k\hv_k^\herm \right ) \wv_k
\end{eqnarray}
Letting $\Sigmam'_z(\lambdav) = \lambda_0 \Id + \sum_{\ell=1}^L \lambda_\ell \Phim_\ell$, we obtain the Lagrangian dual objective function:
\begin{equation} \label{eq:dual-obj}
\Gc(\pv,\lambdav) = \min_{P,\{\wv_k\}} \Lc(P,\{\wv_k\}, \pv, \lambdav)
\end{equation}
It is obvious that $\Gc = -\infty$ if $1 - \lambda_0 < 0$ or the matrix
\[ \Sigmam'_z(\lambdav) + \sum_{j\neq k} p_j \hv_j\hv_j^\herm - \frac{p_k}{\eta_k} \hv_k\hv_k^\herm \]
is not positive semidefinite. On the other hand, adding the constraint
$0 \leq \lambda_0 \leq 1$ and the positive semidefiniteness,  the dual problem takes on the equivalent form:
\begin{eqnarray} \label{eq:min-pow-dual}
{\rm maximize} && \sum_{k=1}^K p_k - \sum_{\ell=1}^L \lambda_\ell \gamma_\ell \nonumber \\
\mbox{subject to} && \left( \Sigmam'_z(\lambdav) + \sum_{j\neq k} \hv_j \hv_j^\herm p_j \right) \succeq \frac{p_k}{\eta_k} \hv_k \hv_k^\herm, \;\;\; \forall k \nonumber \\
&& 0 \leq \lambda_0 \leq 1, \;\;\; \lambdav \geq 0, \;\;\; \pv \geq 0
\end{eqnarray}
Notice that the solution with respect to $\lambda_0$ is trivially obtained by letting $\lambda_0 = 1$.
Hence, we shall replace $\lambda_0$ by 1 in the following. As shown in \cite[Lemma 1]{Yu-Lan-TSP07}, the semidefinite constraints can be rewritten in terms of uplink SINRs as
\[ \SINR_k^{\rm ul} = p_k \hv^\herm_k \left( \Sigmam'_z(\lambdav) + \sum_{j\neq k} p_j \hv_j \hv_j^\herm \right)^{-1} \hv_k \leq \eta_k \]
where the $k$-th uplink SINR is the SINR at the output of a linear MMSE receiver defined by the beamforming vector
\[ \widehat{\wv}_k = \left [ \Sigmam'_z(\lambdav) + \sum_{k=1}^K p_j \hv_j \hv_j^\herm \right ]^{-1} \hv_k \]
Finally, since the SINR constraints must be attained with equality, they can be reversed while turning the
maximization with respect to $\lambdav$ into a minimization, so that the dual problem is given in the desired form
(\ref{eq:min-pow-dual-final}).

\section{Proof of Theorem \ref{thm:feasible-rank1}} \label{appen:feasi-rank1}

Let $\{\Am_k^*\}$ be a solution of the convex relaxation of (\ref{eq:wsrm-zfbf-nozf}) (i.e., after neglecting the rank-1 constraints).
This can be reformulated with respect to the auxiliary ``slack'' variables $\Xim$ as follows:
\begin{eqnarray} \label{eq:wsrm-zfbf1-cacazzi}
\mbox{maximize} &  & \sum_{k=1}^K W_k \log\left (1 +  \mu_k(\Xim) \right ) \nonumber \\
\mbox{subject to} &  & \Xim \succeq 0, \;\;\;  \sum_{k=1}^K [ \Xim ]_{k,\ell} \leq \gamma_\ell, \;\;\; \forall \ell
\end{eqnarray}
where $\mu_k(\Xim)$ is the solution of the auxiliary problem:
\begin{eqnarray}  \label{auxiliary-projected}
\mu(\Xim) & = & \max_{\Am_k \succeq 0} \; [\Am_k]_{1,1} \nonumber \\
\mbox{subject to} & & \trace ( \Am_k \widetilde{\Phim}_\ell) \leq [ \Xim ]_{k,\ell} , \;\;\; \forall \ell
\end{eqnarray}
where $\widetilde{\Phim}_\ell$ are defined as in (\ref{eq:wsrm-zfbf-nozf}).

Problem (\ref{auxiliary-projected}) is a special case of the problem
\begin{eqnarray} \label{eq:max-cov-mat}
\mbox{maximize} && \uv^\herm \Am \uv \nonumber \\
\mbox{subject to} && \Am \succeq 0, \;\;\; \trace ( \Psim_\ell \Am ) \leq \eta_\ell, \;\;\; \forall \; \ell
\end{eqnarray}
where $\uv$, $\{\Psim_\ell \succeq 0\}$ and $\eta_\ell > 0$ are given vector, matrices and constants, respectively. Let $\Am^\star$ denote a solution of (\ref{eq:max-cov-mat}) and assume that the problem
is bounded.\footnote{Since we always consider a sum-power constraint, in our case the problem is always bounded.}
Let $\av^\star$ be the vector solution to the problem
\begin{eqnarray} \label{eq:max-vec}
\mbox{maximize} && \Re \{ \uv^\herm \av \} \nonumber \\
\mbox{subject to} && \av^\herm \Psim_\ell \av \leq \eta_\ell, \;\;\; \forall \; \ell.
\end{eqnarray}
In \cite[Lemma 1]{Wiesel-Eldar-Shamai-TSP08} it is shown that $|\uv^\herm \av^\star|^2 = \uv^\herm \Am^\star \uv$. In turns, thanks to the chain of equivalent problems given above, this fact implies that a rank-1 solution for (\ref{eq:wsrm-zfbf1}) can be found from the solution of the convex relaxation by using (\ref{eq:socp}). Here, we are interested to show that {\em any} feasible point $\widetilde{\Tm}_k = \Um_k \widetilde{\Am}_k \Um_k^\herm$ of the convex relaxation problem can be mapped into a rank-1 feasible point without decreasing the value of the objective function (weighted rate sum).

This is proved if we show that for any feasible point $\widetilde{\Am}$ of (\ref{eq:max-cov-mat}) there exists a vector $\widetilde{\av}$ such that $\widetilde{\av}\widetilde{\av}^\herm$ is also feasible for (\ref{eq:max-cov-mat}) and $|\uv^\herm \widetilde{\av}^\star|^2 \geq \uv^\herm \widetilde{\Am} \uv$.
Furthermore, $\widetilde{\av}$ can be found by solving a SOCP of the type of (\ref{eq:max-vec}).


Define
\begin{equation}  \label{ratio-def}
\alpha = \frac{\uv^\herm \widetilde{\Am} \uv}{\uv^\herm \Am^\star \uv} \leq 1
\end{equation}
and let $\widehat{\av} = \sqrt{\alpha} \av^\star$. Then, we obtain
\[ |\uv^\herm \widehat{\av} |^2 = \alpha |\uv^\herm \av^\star|^2 = \uv^\herm \widetilde{\Am} \uv, \]
and
\[ \widehat{\av}^\herm \Psim_\ell \widehat{\av} = \alpha {\av^\star}^\herm \Psim_\ell \av^\star \leq \eta_\ell. \]
Hence $\widehat{\av}$ is a feasible vector achieving the same objective function of
$\widetilde{\Am}$ (this shows existence). Now, consider the SOCP
\begin{eqnarray} \label{eq:socp-feasi}
\mbox{maximize} && \Re \{ \uv^\herm \av \} \nonumber \\
\mbox{subject to} && \av^\herm \Psim_\ell \av \leq \trace ( \Psim_\ell \widetilde{\Am} ), \;\;\; \forall \; \ell
\end{eqnarray}
Letting $\widetilde{\av}$ denote the solution
of (\ref{eq:socp-feasi}), since by construction $\widehat{\av}$ is a feasible point of (\ref{eq:socp-feasi}), we have
$|\uv^\herm \widetilde{\av}|^2 \geq |\uv^\herm \widehat{\av} |^2 = \uv^\herm \widetilde{\Am} \uv$ which is what we wanted to show.

\bibliographystyle{IEEEtran}
\bibliography{manuscript_rev}

\newpage
\clearpage

\newpage

\newpage

\end{document}